\documentclass[aps,prb,twocolumn,amsmath]{revtex4}
\usepackage[dvips]{color}

\definecolor{g-blue}{rgb}{0.83,0.95,1}
\definecolor{g-yellow}{rgb}{1,1,0.7}
\definecolor{g-green}{rgb}{0.9,1,0.9}
\definecolor{green}{rgb}{0,0.6,0}
\definecolor{cyan}{rgb}{0,0.7,0.7}
\definecolor{black}{rgb}{0,0,0}
\definecolor{grey}{rgb}{0.4 ,0.4 ,0.4 }

\def\white#1{\textcolor{white}{#1}}
\usepackage{bm}
\usepackage{amssymb}
\usepackage{amsfonts}
\usepackage{amsmath}
 \usepackage{graphicx}
  \usepackage{amsmath,bm,epsfig}
\def \ed {\end{document}}
\def\Fbox#1{\vskip1ex\hbox to 8.5cm{\hfil\fboxsep0.3cm\fbox{%
  \parbox{8.0cm}{#1}}\hfil}\vskip1ex\noindent}  

\newcommand{\eq}[1]{(\ref{#1})}
\newcommand{\Eq}[1]{Eq.\,(\ref{#1})}
\newcommand{\Eqs}[1]{Eqs.\,(\ref{#1})}
\newcommand{\Fig}[1]{Fig.\,\ref{#1}}
\newcommand{\Figs}[1]{Figs.\,\ref{#1}}
\newcommand{\Ref}[1]{Ref.\,\cite{#1}}
\newcommand{\Refs}[1]{Refs.\,\cite{#1}}

\def\be{\begin{equation}}\def\ee{\end{equation}}
\def\bea{\begin{eqnarray}}\def\eea{\end{eqnarray}}
\def\bse{\begin{subequations}}\def\ese{\end{subequations}}
\newcommand{\BE}[1] {\begin{equation}\label{#1}}
\newcommand{\BEA}[1]{\begin{eqnarray}\label{#1}}
\newcommand{\BSE}[1]{\begin{subequations}\label{#1}}

\let \nn  \nonumber

\let \= \equiv  \let\*\cdot \let\~\widetilde \let\^\widehat \let\-\overline
\let\p\partial

  \def\1{\bm1} 

\def\<{\left\langle}    \def\>{\right\rangle}
\def\({\left(}          \def\){\right)}
 \def \[ {\left [} \def \] {\right ]}

\renewcommand{\a}{\alpha}\newcommand{\g}{\gamma}
\renewcommand{\d}{\delta}
\newcommand{\ve}{\varepsilon}
\renewcommand{\o}{\omega} \renewcommand{\O}{\Omega}

\def\r{\rho}

\newcommand{\B}[1]{{\bm{#1}}}
\newcommand{\C}[1]{{\mathcal{#1}}}    

\renewcommand{\sb}[1]{_{\text {#1}}}  
\renewcommand{\sp}[1]{^{\text {#1}}}  
\def\Sb#1{_{\scriptscriptstyle\rm{#1}}}
\def\He4 {$^4$He~}

\begin{document}

\title{
Turbulent statistics and intermittency enhancement in coflowing superfluid $^4$He  }
\author{L. Biferale$^1$, D. Khomenko$^{2,3}$, V. L'vov$^2$, A. Pomyalov$^2$, I. Procaccia$^2$ and G. Sahoo$^{4}$}
\affiliation{$^1$Dept. of Physics and INFN, University of Rome, Tor Vergata,
Roma, Italy\\
$^2$Dept. of Chemical Physics, Weizmann Institute of Science, Rehovot, Israel\\
$^3$Laboratoire de physique th\'{e}orique, D\'{e}partement de physique de l'ENS, \'{E}cole normale sup\'{e}rieure, PSL Research University, Sorbonne Universit\'{e}s, UPMC Univ. Paris 06, CNRS, 75005 Paris, France.\\
$^4$Dept. of Mathematics and Statistics and Dept. of Physics, University of Helsinki, Finland}
\begin{abstract}
The large scale turbulent statistics of mechanically driven superfluid $^4$He was shown experimentally to follow the classical counterpart. In this paper we use direct numerical simulations to study the whole range of scales in a  range of temperatures $T \in[1.3,2.1]$~K. The numerics employ self-consistent and non-linearly coupled normal and superfluid components. The main results are that (i) the velocity fluctuations of normal and super components are well-correlated in the inertial range of scales, but decorrelate at small scales. (ii) The energy transfer by mutual friction between components is particulary efficient in the temperature range between 1.8~K and 2~K, leading to enhancement of small scales intermittency for these temperatures. (iii) At low $T$ and close to $T_{\lambda}$ the scaling properties of the energy spectra and structure functions of the two components are approaching those of classical hydrodynamic turbulence.
\end{abstract}

\maketitle
\section{\label{s:intro}Introduction}

Superfluid $^4$He below the transition temperature $T_{\lambda}= 2.17$~K may be viewed as a two-fluid
system \cite{11,12,Rev1,HV,BK} consisting of a normal fluid with very low kinematic viscosity $\nu \sb n(T)$
and an inviscid superfluid component. The contributions of the components  are defined by their densities, $\rho_n(T), \rho \sb s(T)$,  constituting together the density of superfluid $^4$He: $ \rho\sb {He}=\rho\sb n(T)+\rho \sb s(T)$. Each component moves with its own velocity $u
\sb n(\bm r, t), u\sb s(\bm r, t)$. Due to quantum mechanical restrictions, the circulation in the superfluid component is confined to thin vortex lines and quantized to multiples of circulation quantum $\kappa=h/m \approx 10^{-3}$ cm$^2$/s, where $h$ is the Plank constant and $m$ denotes the mass of a $^4$He atom. Turbulence in the superfluid component is accompanied by the creation of a dense disordered tangle of these vortex lines with a typical intervortex distance $\ell$.

It is commonly accepted that the statistical properties of the large scale fluctuation in turbulent superfluid He conform with those of classical fluids when forced by mechanical means.  Examples are rotating containers or flows behind a grid.
The  mean velocities of the normal and superfluid components in such driven superfluid $^4$He appear to coincide\cite{Tabeling}. Numerous laboratory and theoretical
studies showed that under these conditions the mutual friction between the normal- and superfluid
components couples also their fluctuations: $u\sb n(\bm r, t)\approx u\sb s(\bm r, t)$ almost at all scales and the resulting
turbulent energy spectra of the mechanically driven superfluid turbulence are close to those of the
classical hydrodynamic turbulence\cite{Roche1,Roche2,Salort11,Rev1,Rev2,Rev3}.

One of the important aspects of the turbulent statistics in classical turbulence is the intermittency of  the velocity fluctuations. This intermittency  results in corrections to the dimensionally derived energy spectra and  structure functions. This subject is thoroughly studied in the classical case, but much less so in the context of  superfluid $^4$He. The experiments \cite{Roche1,Roche2,Salort11}, conducted mostly at low temperatures  and close to $T_{\lambda}$, did not find deviations from the turbulent statistics of classical flows. A very recent experimental study \cite{Roche-new} of turbulence in the wake of a disc was conducted in a wide range of temperatures;  it also did not find any temperature dependence of the scaling exponent of the second order structure function. Preliminary results of the ongoing study \cite{WeiEmilGrid} of turbulence behind a grid indicated a temperature dependence of higher-order structure functions scaling.

Numerical simulations of homogeneous isotropic superfluid turbulence using  different methods indicated that turbulence statistics in $^4$He depend on the temperature.  The important aspect is the relative density of the normal and super components. At temperature close to the $T_\lambda$ and also for $T\lesssim 1.6$K, where one or the other component dominate, the scaling exponents of structure functions are close to those of classical turbulence. In the range of temperatures where the densities of the components are similar, the statistics change. The shell-model study\cite{He4} found larger intermittency corrections compared to classical turbulence in these  conditions. It was conjectured that the effect is related to the energy exchange between the normal and superfluid components and   the additional dissipation due to mutual friction between components.
Recently these findings were questioned in another shell-model study \cite{Shukla}, where intermittency was found to be suppressed for the same conditions or even absent in a certain temperature range.
This conjecture appears to disagree with the results of the Gross-Pitaevkii simulations\cite{GPEgrid} of grid turbulence. There enhanced intermittency is found in the zero-temperature limit. In light of these conflicting results is appears worthwhile to investigate further these issues.

We present here results of  Direct Numerical Simulations (DNS) of mechanically driven superfluid $^4$He. We study  energy spectra and structure functions of both components in a wide range of temperatures, using  typical parameters\cite{DB98} for $^4$He. A  non-monotonic temperature dependence of the apparent scaling exponents of the energy spectra and the structure functions is found. The exponents are close to  their classical counterparts at low temperatures and close to  $T_\lambda$. In the intermediate temperature range $1.8\lesssim T\lesssim 2$~K, where the densities of the components are similar $\rho\sb s \sim \rho\sb n$, the scaling properties  significantly deviate  from their classical values. The difference in properties can be attributed to the degree of dynamical correlations between the fluctuations of the two components. The normal and superfluid velocity fluctuations  appear correlated at low and high temperatures, but almost uncorrelated in the intermediate temperature range. Then the small scale intermittency measured by the velocity flatness  is found to strongly exceed the classical values. The  analysis of the energy balance at different scales revealed the role of the dissipation by mutual friction in intermittency enhancement.

\section{\label{s:He3} Statistics  of turbulence in co-flowing $\bm ^4$He : analytical description}

\subsection{\label{ss:HVBK} Gradually damped HVBK-equations for   superfluid $^3$He-B turbulence}
Following \Ref{DNS-He3} we describe large scale turbulence in superfluid $^4$He   by the gradually-damped version\cite{He4} of the coarse-grained  Hall-Vinen \cite{11}-Bekarevich-Khalatnikov\cite{12} (HVBK)  equations. It  has a form of two Navier-Stokes equations  for $\B u\sb n(\B r,t)$ and $\B u\sb s(\B r,t)$:
\begin{subequations}\label{NSE} \begin{eqnarray}   \label{NSEs} 
 && \hskip -1.3cm \frac{\p \,\B u\sb s}{\p t}+  (\B u\sb s\* \B\nabla) \B u\sb s
 - \frac 1{\rho\sb s }\B \nabla p\sb s  =\nu\sb s\,  \Delta \B u\sb s   + \B f \sb {ns}+\B \varphi\sb s\,, 
 \\  \label{NSEn}
&& \hskip -1.3cm \frac{\p \,\B u\sb n}{\p t}+(\B u\sb n \* \B \nabla)\B u\sb n
- \frac 1{\rho\sb n }\B \nabla p\sb n = \nu\sb n\,  \Delta \B u\sb n -\frac{\rho\sb s}{\rho\sb n}\B f \sb {ns}+\B \varphi\sb n\, ,\\ \nn
&& \hskip -1.3cm  p\sb n =\frac{\rho\sb n}{\rho }[p+\frac{\rho\sb s}2|\B u\sb s-\B u\sb  n|^2]\, ,
   p\sb s =  \frac{\rho\sb s}{\rho }[p-\frac{\rho\sb n}2|\B u\sb s-\B u\sb n|^2]\, ,\\ \label{1e}
   \B f\sb {ns}&\simeq& \Omega  \,(\B  u \sb n-\B  u \sb s ) \,,
   \quad \Omega = \a(T)\, \Omega\Sb T\,,
\end{eqnarray}\end{subequations} 
stirred by a random force $\B \varphi(\B r,t)$ and coupled by the mutual friction force $\B f\sb{ns}$ in approximated
form\cite{LNV}.\Eq{1e}.  It involves the temperature
dependent dimensionless dissipative mutual friction parameter $\alpha(T)$ and rms
superfluid turbulent vorticity $\Omega\Sb T$.
In isotropic turbulence
\begin{equation}\label{ot}
 \Omega\Sb T ^2(t) \= \frac12\< |\B \o(\bm r,t)|^2\>_{\bm r}\approx  \int  k^2  E\sb s(k,t)d k\,,
\end{equation}
 where $E\sb s(k,t)$ is the one-dimensional (1D) energy
spectrum, normalized such that the total energy density per
unit mass $\C E\sb s(t) =\int  E\sb s (k,t)\, d k$.  Here the turbulent vorticity $\Omega\Sb T(t)$ is defined self-consistently from the instantaneous energy spectrum  $E\sb s(k,t)$.
Other parameters include the pressures  $p\sb n$,  $p\sb s$ of the normal and the superfluid components,
the He density $\rho\equiv  \rho\sb s+\rho\sb n$ and the kinematic viscosity  of normal fluid component $\nu\sb n$. The dissipative term with the Vinen's effective superfluid viscosity  $\nu\sb s$  was added\cite{He4} to account for the energy dissipation at the intervortex scale $\ell$ due to vortex reconnections and similar effects.

Generally speaking,   \Eqs{NSEs}  and \eqref{NSEn} involve also   contributions of a reactive (dimensionless)   mutual friction  parameter $\alpha'$, that renormalizes  their nonlinear terms. For example, in \Eq{NSEs} $ (\B u\sb s\* \B
\nabla) \B u\sb s  \Rightarrow  (1-\alpha')(\B u\sb s\* \B
\nabla) \B u\sb s $.  However,  in the studied range of temperatures $|\alpha'| \lesssim 0.02 \ll 1$ [see column (6) in Tab.\,\ref{t:1}] and this   renormalization can be peacefully ignored. For similar reasons we neglected all other $\alpha'$-related term in \Eqs{NSE}.


 \subsection{\label{ss:defs}  Statistical description of   space-homogeneous, isotropic turbulence of superfluid $^3$He}
\subsubsection{\label{ss:defs1}Definition of 1-D energy spectra and cross-correlations}
Traditionally the energy distribution over scales in a space-homogeneous, isotropic case is described by  one-dimensional (1D) energy  spectra  of the normal  and superfluid components, $E\sb n(k)$ and $E\sb s(k)$:
 \begin{subequations}\label{corr}\begin{equation}
 E\sb{n}(k)= \frac{k^2}{2\pi^2}F\sb{nn}(k)\,, \quad   E\sb{s}(k)= \frac{k^2}{2\pi^2}F\sb{ss}(k)\,,\\ \label{def1a}
  \end{equation}
 defined in terms of the three-dimensional spectra $F\sb {nn}(k)$ and $F\sb {ss}$:
\begin{eqnarray}\label{corr-nn}
&&\< \~{\B u}\sb n(\B k,t)\*\~{\B u}  \sb n(\B q ,t)  \>= (2\pi)^3 F\sb{nn}(\B k)\, \d(\B k+\B q )\,,~~~ \\
\label{corr-ss}
&& \< \~{\B u}\sb s(\B k,t)\*\~{\B u}  \sb s(\B q ,t)  \>=(2\pi)^3 F\sb{ss}(\B k)\, \d(\B k+\B q )\ .
\end{eqnarray}\end{subequations}
 Here $ \~{\B u}\sb n(\B k,t)$ and $ \~{\B u}\sb s(\B k,t)$ are $\B k$-Fourier transforms of the velocity fields $ {\B u}\sb n(\B r,t)$ and $ {\B u}\sb s(\B r,t)$. Delta-function $\delta(\B k+\B q )$  is a consequence of space homogeneity.

 Similarly to the energy spectra, we define a simultaneous cross-correlation function:
 \begin{subequations}\label{Xcorr}
 \begin{eqnarray}\label{XcorrA}
  E\sb{ns}(k)&\=& \frac{k^2}{2\pi^2}F\sb{ns}(k)\,, \\ \label{XcorrB}
 \< \~{\B u}\sb n(\B k,t)\*\~{\B u}  \sb s(\B q ,t)  \>&=& (2\pi)^3 F\sb{ns}(\B k)\, \d(\B k+\B q )\ .
 \end{eqnarray}\end{subequations}
and a third-order correlation function
 \begin{eqnarray}\nn  && \< \~{  u}^{\,\xi} \sb s (\B k,t)\, \~{ u}^{\, \beta} \sb s(\B q,t)\, \~{u}^{\, \gamma}  \sb s(\B p ,t)  \> \\ \label{Corr-sss}
&=& (2\pi)^3 F ^{\xi \beta \gamma }\sb{sss}(\B k,\B q, \B p)\, \d(\B k+\B q +\B p)\ .
\end{eqnarray}

\subsubsection{\label{ss:defs2}Energy balance equation in the $k$-representation}

The balance equations\cite{LP-QFS} for superfluid and normal fluid  energy spectra,  $E\sb s(k,t)$ and  $E\sb n (k,t)$ in stationary case read:
  \begin{subequations}\label{BAL} \begin{eqnarray}\label{BALas}
  &&   \frac{\partial \ve\sb s(k)}{\partial k }+ \mbox{D}_{\rm s,\nu}(k) + \mbox{D}_{\rm s,\alpha}(k)=0\,, \\ \label{BALan}
   &&   \frac{\partial \ve\sb n(k)}{\partial k }+ \mbox{D}_{\rm n,\nu}(k) + \mbox{D}_{\rm n,\alpha}(k)=0\,, \\ \label{BALbs}
 && \mbox{D}_{\rm s,\nu}(k) =  2\, \nu\sb s k^2  E\sb s (k)\,, \ \mbox{D}_{\rm n,\nu} =  2\, \nu\sb n k^2  E\sb n (k)\,, \\ \label{BALms}
 &&   \mbox{D}_{\rm s,\alpha}(k)  = 2\, \Omega  \big[ E\sb{s}(k)-E\sb {ns} (k)  \big]\,,\\\label{BALmn}
 && \mbox{D}_{\rm n,\alpha}(k)  = 2\, \Omega\frac{\rho\sb s}{\rho\sb n}  \big[ E\sb{n}(k)-E\sb {ns} (k)  \big]\ .
   \end{eqnarray} \end{subequations}
 Here terms  D$_{\rm s,\nu}$ and  D$_{\rm n,\nu}$ describe viscous energy dissipation. The terms  D$_{\rm s,\alpha}$  and  D$_{\rm n ,\alpha}$ are responsible for the energy dissipation by mutual friction with characteristic frequency $\Omega $  given by \Eqs{1e} and  \eqref{ot}.

To obtain a simple form \Eqs{BALms} and \eqref{BALmn} we, following \Ref{LNV}, accounted for the fact that $\Omega(t)$ is dominated by the motions of smallest scales (about intervortex distance $\ell$), while $E(k,t)$ - by the  $1/k$ scale. This allows us to neglect their correlation in time and to replace $\langle \Omega(t) E(k,t)\rangle$ by a product of $\langle \Omega(t)\rangle_t=\Omega$ and $\langle E(k,t)\rangle_t=E(k)$.

First terms of the balance equations are  related to the energy transfer over scales.
The energy transfer term Tr$(k)$ in \Eqs{BAL} (in which we omit here subscripts $``\sb s"$ and $``\sb n"$) originates from the nonlinear terms in the HVBK \Eqs{NSEs} and has the same form as in classical hydrodynamic turbulence (see, e.g. \Refs{LP-95,LP-2}):
 \begin{subequations}\label{flux}  \begin{eqnarray}\nn
 \mbox{Tr}(\B k)&=& 2\, \mbox{Re}\Big\{\int V^{\xi\beta\gamma}(\B k,\B q,\B p)\, F^{\xi\beta\gamma}(\B k,\B q,\B p)\\ \label{genA}
 && \times \delta(\B k+\B q+\B p)\frac{d^3 q \, d^3 p}{(2\pi)^6} \, \Big\}\,, \\ \nn
 V^{\xi\beta\gamma}(\B k,\B q,\B p)&=& i  \Big ( \delta _{\xi \xi'}- \frac{ k^\xi k^{\xi'}}{k^2}   \Big )\\ \label{genB}
  && \times \Big( k^\beta \delta_{\xi ' \gamma} + k^\gamma \delta _{\xi' \beta} \Big )\ .
 \end{eqnarray}
  Importantly, Tr$(k)$ preserves total turbulent kinetic energy: $\displaystyle \int_0 ^\infty\mbox{Tr}(k')dk'=0$ and therefore can be written in the 1D divergent  form:
\begin{equation}\label{BALb}
\mbox{Tr}(k)= \frac{\partial\,  \ve (k)}{d k}\,,
\end{equation}\end{subequations}
where $\ve(k)$ is the energy flux over scales.


\begin{figure*}
 \begin{tabular}{ccc}
(a) & (b) & (c) \\
\white{.}\hskip -.4 cm  \includegraphics[scale=0.35]{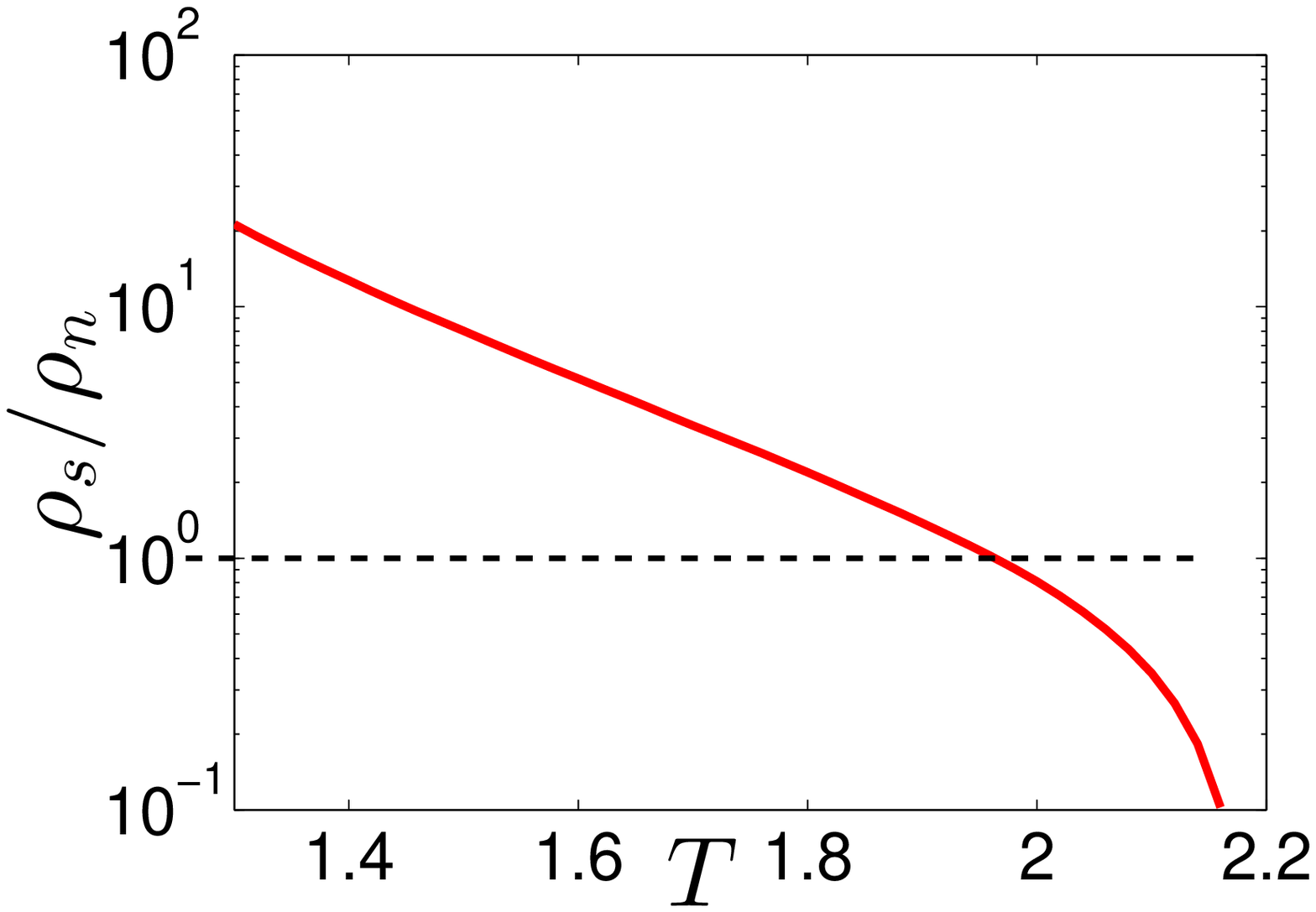}& \hskip -.3 cm
  \includegraphics[scale=0.35 ]{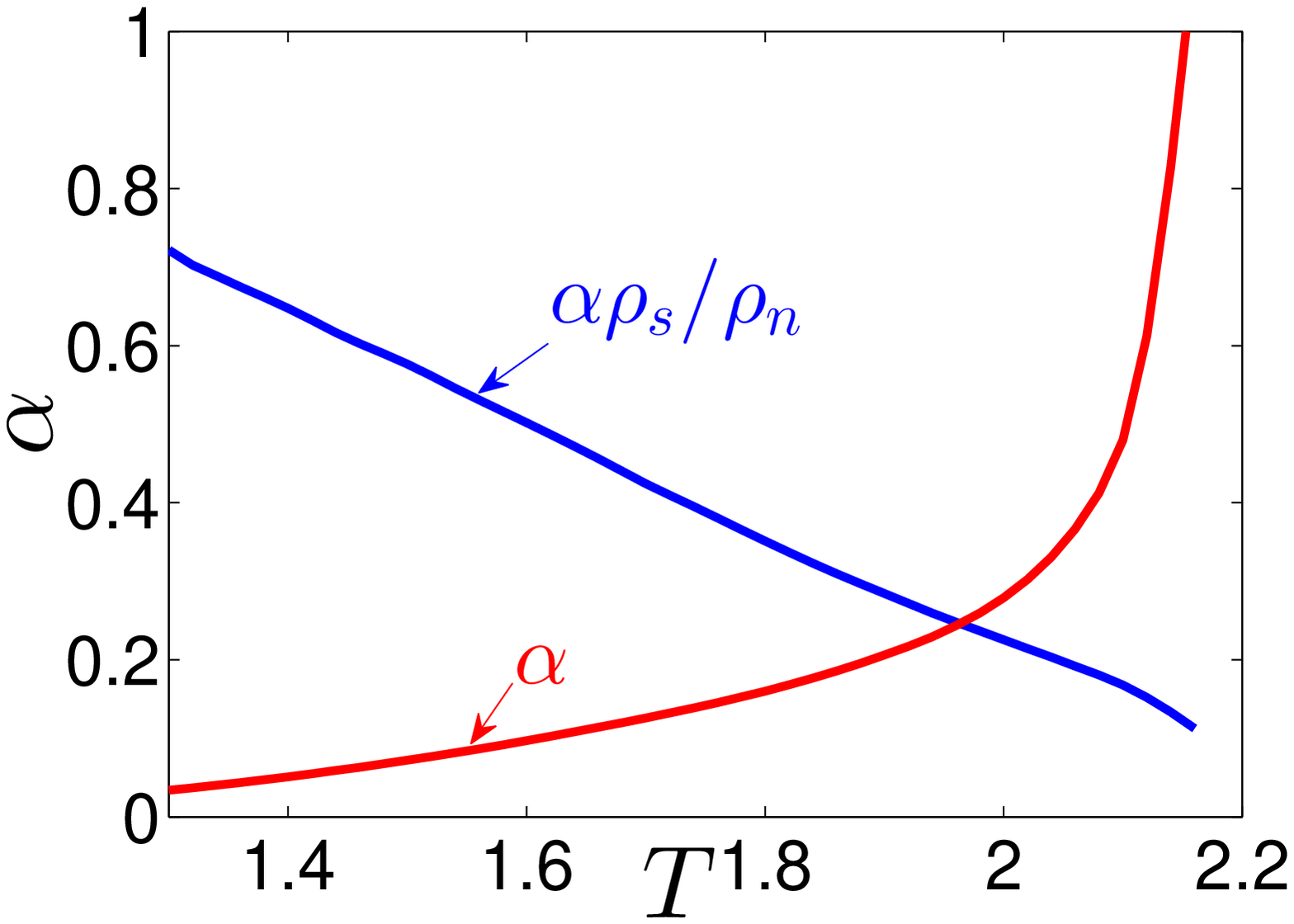}& \hskip -.3 cm
   \includegraphics[scale=0.35]{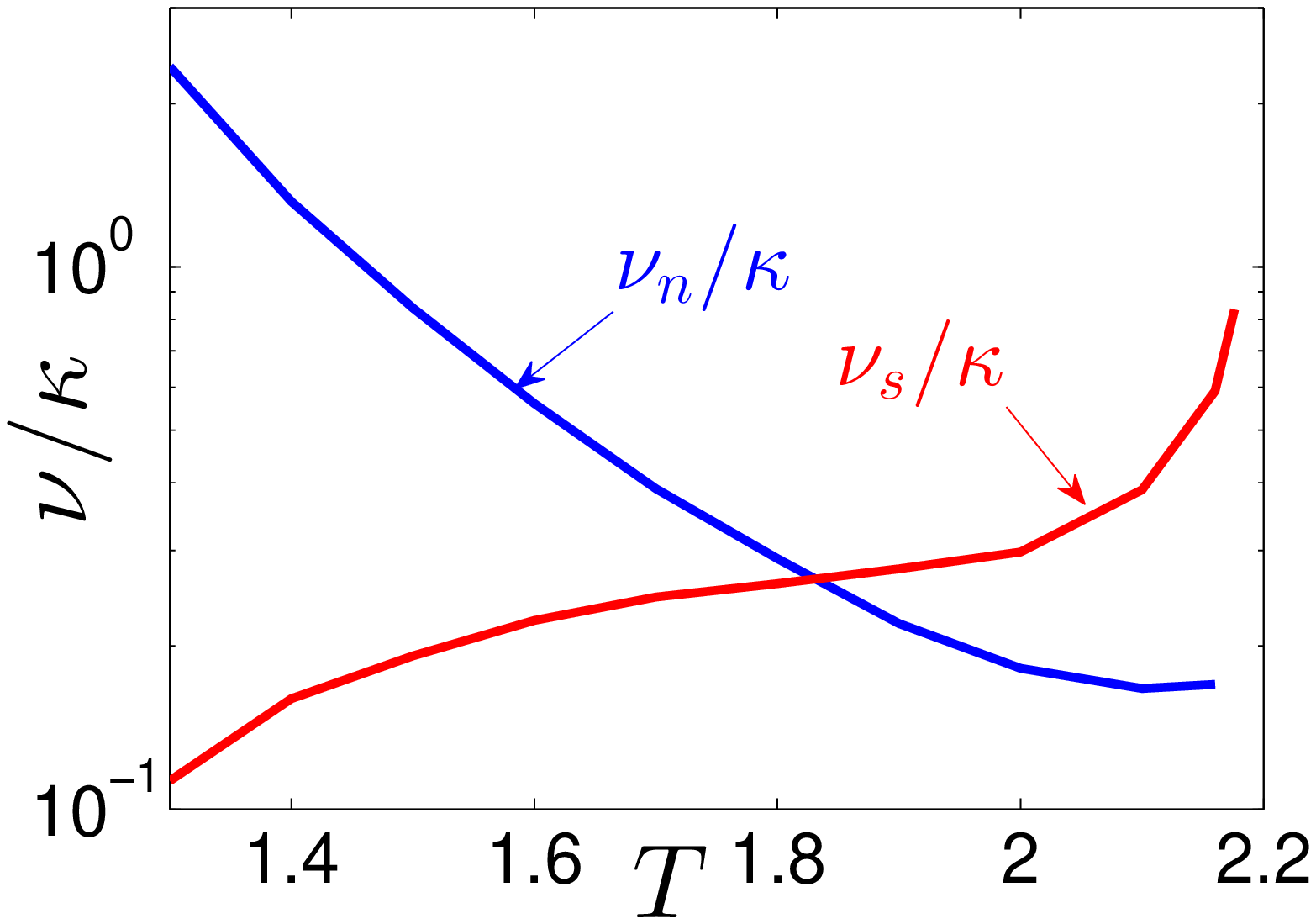}
 \\
   \end{tabular}
\caption{\label{F:1} Color online. Temperature dependence of $^4$He parameters used in the simulations of HVBK \Eqs{NSE}:   ratio of the superfluid and normal fluid densities [Panel (a)], mutual friction parameters $\alpha$ in the superfluid \Eq{NSEs} and $\alpha\rho\sb s/\rho \sb n$ in the normal fluid \Eq{NSEn} [Panel (b)] and (effective) kinematic viscosities $\nu\sb s$ and $\nu\sb n$ [Panel (c)]. }
\end{figure*}

\begin{table*}[t]
\caption{\label{t:1}  Parameters of simulations by columns:
  (\# 1) Temperature; (\# 2) Ratio of the normal and superfluid densities, $\rho\sb n / \rho \sb s$;
  (\# 3 and \#4)  the kinematic viscosity of the superfluid and normal fluid components $\nu\sb s $ and  $\nu\sb s $;
   (\# 5 and \#6) Parameters of the mutual friction $\alpha $ and $\alpha'$;
   (\# 7 and \#8) the root mean square (rms) of the superfluid and normal fluid turbulent velocity fluctuations $v\sp s\Sb T$ and $v\sp n\Sb T$;
    (\# 9 and \#10)   the  rms  of the superfluid and normal fluid vorticity $\Omega\sp s\Sb T$ and $\Omega\sp n\Sb T$;
   (\# 11 and \#12)  the Taylor-microscale Reynolds number of  the superfluid and normal fluid components Re$\sp s_\lambda$ and Re$\sp n_\lambda$,   where $\displaystyle \lambda = 2\pi u\Sb T/ \Omega\Sb T$ is the Taylor  microscale;\\
 The temperature dependence of $\rho\sb s/ \rho \sb n$, $\nu\sb n, \alpha$ and $\alpha'$  are given according to \Ref{DB98} and $\nu\sb s$ according to \Ref{PRB}. In all simulations: the number of collocation points along each axis is $N=1024$;
 the size of the periodic box is $L=2\pi$  the range of forced wavenumbers $k^{\tilde \varphi}=[0.5 , 1.5]$.  }
\begin{tabular*}{\linewidth}{@{\extracolsep{\fill} } |     c    c |  c   c | c  c|  c  c|   c   c | c    c |}
    \hline
    \hline
 1 &2&3&4&5&6&7&8&9&10&11&12 \\ \hline
   ~~$T$~~ &$\displaystyle \frac{\rho\sb s }{ \rho \sb n}$ & $\nu\sb s$ &$\nu \sb n $ & ~~$\alpha$~~ & ~$\alpha'~$~~ & $u\sp n \Sb T=$ & $u\sp s \Sb T=$ & $\Omega\sp n \Sb T=$ & $\Omega\sp s \Sb T=$ & Re$\sp n_\lambda=$  &Re$\sp s_\lambda=$    \\
   K & &$\times 10^4$ & $\times 10^4$  & $\times 10 $ & $\times 10$&$ \sqrt {\< {u\sp n}^2 \> } $ &$ \sqrt {\< {u\sp s}^2 \> } $ &$ \sqrt {\< {\omega\sp n}^2 \>/2 } $&$ \sqrt {\< {\omega\sp s}^2 \>/2 } $ & $u\sp n\Sb T \lambda\sb n /\nu\sp n$&$u\sp s\Sb T \lambda\sp s /\nu\sb s$  \\
    \hline \hline
  1.3 & 20.0& 5.0 & 117.0 & 0.34 & 0.14& 4.3 & 4.5 & 33 & 62 &222&2865   \\
  1.6 &  5.0& 5.0 & 13.5& 0.97 &  0.16 & 4.2 & 4.2 & 45 & 57 & 1285 & 2770    \\
  1.8 &2.17 & 5.0 & 6.1&  1.6 &  0.08 & 3.6 & 3.6 & 31 & 37 & 2910 & 3151    \\ \hline
 1.9 & 1.35 & 5.0 & 4.0 & 2.06&  0.08 & 3.7 & 3.7 & 30 & 30 & 4042 & 3182    \\
  1.9 & 1.35 & 6.3 &5.0& 2.06 &  0.08 & 3.4 & 3.5 & 31 & 31 & 4182 & 3306   \\ \hline
   2.0 & 0.81 & 5.0 & 3.0 &  2.79 &  0.12& 3.5 & 3.5 & 24 & 21 & 4448 & 3020   \\
   2.0 & 0.81& 8.6 &5.0    &  2.79& 0.12 & 3.3 & 3.3 & 21 & 18 & 7639 & 5396    \\ \hline
  2.1 & 0.35& 5.0 & 2.0&  4.81 & $-0.24$  & 3.6 & 3.5 & 35 & 25 & 3318 & 17605   \\
   2.1 &0.35& 1.25 &5.0&  4.81 & $-0.24$& 4.3 & 4.2 & 70 & 52 & 5738 & 3083   \\ \hline \hline

  \end{tabular*}

\end{table*}
\subsection{Generalized Kolmogorov's $\dfrac 45$-law for superfluid turbulence}

One of the best known results in the statistical theory of the homogeneous, stationary, isotropic, fully developed
 hydrodynamic  turbulence is the Kolmogorov's  ``four-fifth law"\cite{45law}, which relates the third order structure function
 \begin{subequations}\label{Kolm}
 \begin{equation}\label{KolmA}
 S_{3}\sp l  (R) =\<\big [\delta u\sp l (\B r, \B R )\big ]^3 \>
  \end{equation}
  of the longitudinal velocity differences
   \begin{eqnarray}\label{KolmB}
  \delta u\sp l (\B r, \B R )&\=& \delta \B u  (\B r, \B R )\cdot \B R / R\,, \\ \delta \B u  (\B r, \B R ) &\=&[\B u (\B r + \B R)- \B u (\B r ) ]\,,
  \label{KolmC}
   \end{eqnarray}\end{subequations}
  to the rate of energy dissipation $\ve $. Note that  we omitted for shortness the time argument $t$ in notations for the velocity field $\B u (\B r,t)$ and the correlation functions.
In the inertial interval of scales this law reads:
\begin{equation}\label{4/5}
   S_{3,\rm l}(R)   = - \frac 45 \, \ve R \ .
\end{equation}
Formulated in the $\B R$-space, \Eq{4/5} is much simpler than its equivalent \Eqs{flux}  in the $\B k$-representation that relate the third-order correlation function $\B F\sb {sss}(\B k, \B q, \B p)$,  \Eq{Corr-sss}, to the energy flux $\ve(k)$.

 To generalize the $\dfrac 45$-law\,\eqref{4/5} to the case of superfluid turbulence, define  the second   order velocity structure functions of  the normal and superfluid velocity differences\cite{LPP-97}
\begin{subequations}\label{def1}
\begin{eqnarray}\label{def1A}
S\sb{s} (\B R)&=&\< |\delta \B u \sb {s} (\B r, \B R )|^2\>\,,\\  \label{def1B} S\sb{n} (\B R)&=&\< |\delta \B u \sb {n} (\B r, \B R )|^2\>\,,
 \end{eqnarray}
 and triple correlations
 \begin{eqnarray}\label{def1C}
 J\sb{s }^{\alpha,\beta\gamma}(\B R)&=&\< u\sb{s }^\alpha(\B R+ \B r)u\sb{s }^\beta(\B r) u\sb{s }^\gamma(\B r)  \> \,, \\
 J\sb{ n}^{\alpha,\beta\gamma}(\B R)&=&\< u\sb{ n}^\alpha(\B R+ \B r )u\sb{ n}^\beta(\B r ) u\sb{ n}^\gamma(\B r )  \> \ .\label{def1D}
\end{eqnarray}
From  now on  we omit subscripts ``$\sb n$" or ``$\sb s$" in relations that are valid for both normal and superfluid components.  We also assume spatial homogeneity and $\B R \Leftrightarrow -\B R$ symmetry (i.e.  absence of helicity).   In that case \Eqs{def1A} and \eqref{def1B} can be rewritten as follows:
\begin{equation}\label{def1E}
S(\B R)=- 2\, \<   \delta \B u(\B r,\B R) \cdot \B u(\B r) \>\ .
\end{equation}

\end{subequations}

 Similarly, the  equations for  the third-order structure function $\< \delta  u^\alpha (\B r,\B R)\delta  u^\beta (\B r,\B R)\delta  u^\gamma\ (\B r,\B R)\> $ can be expressed via triple correlations\,\eqref{def1C} and \eqref{def1D}.
 In the stationary turbulence the velocity structure function \eqref{def1E} is time-independent.
 Computing  its rate of change  using \Eqs{NSE},   we find
\begin{subequations}\label{9}\begin{equation} \label{9A}
0=\frac{\partial S(\B R)}{2\,\partial t}=-T(\B R) + P(\B R)- D_\nu (\B R)  - D_\alpha (\B R)\ .
\end{equation}
Here the energy transfer term $T(\B R)$ originates from the nonlinear term $(\B u \cdot \B \nabla) \B u$ in \Eqs{NSE}, the energy pumping term $P(\B R)$ -- from  the random driving force $\B \varphi(\B r,t)$ , the dissipation terms  $D_\nu (\B R)$ and   $D_\alpha (\B R)$ -- from the viscous ($\propto \nu$) and the mutual friction ($\propto \alpha$) terms.

Taking into account that due to space homogeneity, the one-point contribution vanishes\,\cite{LPP-97}, the transfer term may be written as:

 \begin{eqnarray}
  \label{9B}
&& T(\B R)= 2\,\< u^\alpha (\B r+ \B R)  \frac{\partial}{\partial r^\beta}  \, u^\beta(\B r) u^\alpha (\B r)  \> \\ \nn
&&\hskip -.5 cm = -2\, \< \, u^\beta(\B r) u^\alpha (\B r)   \frac{\partial}{\partial r^\beta}u^\alpha (\B r+ \B R) \> \\ \nn
&&\hskip -.5 cm = -2\, \< \, u^\beta(\B r) u^\alpha (\B r)   \frac{\partial}{\partial R^\beta}u^\alpha (\B r+ \B R) \>  =-2 \, \frac{\partial  J^{\alpha, \beta \alpha}(\B R)  }{\partial R^\beta} \ .\end{eqnarray}

Here summation over repeated indices is implied.

 The rest of contributions to \eqref{9A} can be found straightforwardly:
  \begin{eqnarray}  \label{9C}
  P(\B R)&=& 2\varepsilon\,, \quad D_\nu(\B R)= \nu \nabla^2 S(\B R)\,,
  \\   \label{9D}
  D_{\alpha, \rm s}(\B R)&=&\Omega [ S\sb{ns}(\B R)- S\sb s(\B R)]\,,\\   \label{9E} D_{\alpha, \rm n}(\B R)  &=&\frac{\rho\sb s}{\rho\sb n}\, \Omega [ S\sb{ns}(\B R)- S\sb n(\B R)]\,,   \\
 S\sb{ns}(\B R)&\=&\< \delta \B u\sb{s}  ( \B r,  \B R) \cdot \delta \B u\sb{n}  ( \B r,  \B R)\>\ .
  \end{eqnarray}  \end{subequations}

Equations\,\eqref{9} can be rewritten in a compact form:
\begin{subequations} \label{res1}
 \begin{eqnarray} \label{res1A}
 T(\B R)&=& -2 \,   \B \nabla \cdot \B J(\B R) = 2\varepsilon - \nu \nabla^2 S(\B R)- D_\alpha (\B R)\,,~~~~~\\  \label{res1B}
 \B J(\B R)&\=& \< \B u (\B r)[\B u(\B r) \cdot \B u(\B r+\B R) ] \> \,,
\end{eqnarray}\end{subequations}
where $D_\alpha (\B R)$ is given by \Eq{9D} or \eq{9E}.
Equations\,\eqref{res1} represent the generalized form of the Kolmogorov's $\dfrac 45$-law for superfluid turbulence.

To test its consistency with the original form, we consider \eqref{res1} in the inertial interval of scales
of the isotropic turbulence without helicity. First, we recall the most general form of $J ^{\alpha,\beta\gamma}$ in that case \cite{LPP-97}:
\begin{subequations}\label{6}\begin{eqnarray}\label{6A}
&&J^{\alpha,\beta\gamma}=a_1(R)[\delta_{\alpha\beta}R^\gamma + \delta_{\alpha\gamma}R^\beta +\delta_{\beta\gamma}R^\alpha]\\ \nn
&&+a_2(R)[\delta_{\alpha\beta}R^\gamma + \delta_{\alpha\gamma}R^\beta -2 \delta_{\beta\gamma}R^\alpha]\\ \nn
&&+ a_3(R)[\delta_{\alpha\beta}R^\gamma + \delta_{\alpha\gamma}R^\beta + \delta_{\beta\gamma}R^\alpha- 5 R^\alpha R^\beta R^\gamma /R^2]\ .
\end{eqnarray}
Incompressibility conditions result in two relations   between  three functions $a_1(R)$, $a_2(R)$ and $a_3(R)$:
\begin{eqnarray}\label{6B}
&& \Big( \frac d {dR}+\frac5R \Big)a_3(R)= \frac 23 \frac d{dR}\big [ a_1 (R)+a_2 (R) \big ]\,,\\  \label{6C}
&& \Big( \frac d {dR}+\frac3R \Big) \big [ 5 a_1 (R)- 4 a_2 (R) \big ]=0\ .
\end{eqnarray}\end{subequations}
 Using \Eqs{6} one can simplify \Eq{9B} for the transfer term:
 \begin{equation}\label{7}
 T(R)= 2\,\frac {\partial }{\partial R^\alpha}\, R^\alpha \big [ 5 a_1(R)+ 2 a_2 (R)\big] \ .
 \end{equation}
 In the inertial interval of scales, [i.e. $D_\alpha=0$ and $D_\nu=0$],  the term $T(R)$ is independent of $R$. This is possible if $a_1$ and $a_2$ are independent of $R$ as well. Then from
\Eqs{6B}, \eqref{6C}, \eqref{7} together with $T(R)=2 \, \varepsilon$ we find:
\begin{equation}\label{8}
a_1=- 2\varepsilon / 45\,,\quad  a_2=-\varepsilon /18, \quad a_3=0\ .
\end{equation}
Together with \Eq{6A} this finally gives:
\begin{equation}\label{10} J^{\alpha,\beta\gamma}= -\frac{\varepsilon}{10}\Big(R^\gamma \delta_{\alpha\beta}+ R^\beta \delta_{\alpha\gamma}- \frac 23 R^\alpha \delta_{\beta\gamma} \Big)\ .
\end{equation}
The longitudinal third-order structure function $S_3\sp l$, \Eqs{Kolm}
can be rewritten as follows:
\begin{equation}\label{11}S_3\sp l(R)=   6 \<u\sp l(\B r+\B R) (u\sp l(\B r))^2 \> =   6 J^{z,zz}(R)\ ,
\end{equation}
where $u\sp l(\B r+\B R)=u(\B r+\B R)\cdot \hat z\, , \hat z= \B R / R$.
Together with \Eqs{10} this gives  the celebrated $\dfrac45$-Kolmogorov's law\,\eqref{4/5}.

\begin{figure*}
 \begin{tabular}{cc}
(a) & (b)   \\
    \includegraphics[scale=0.4 ]{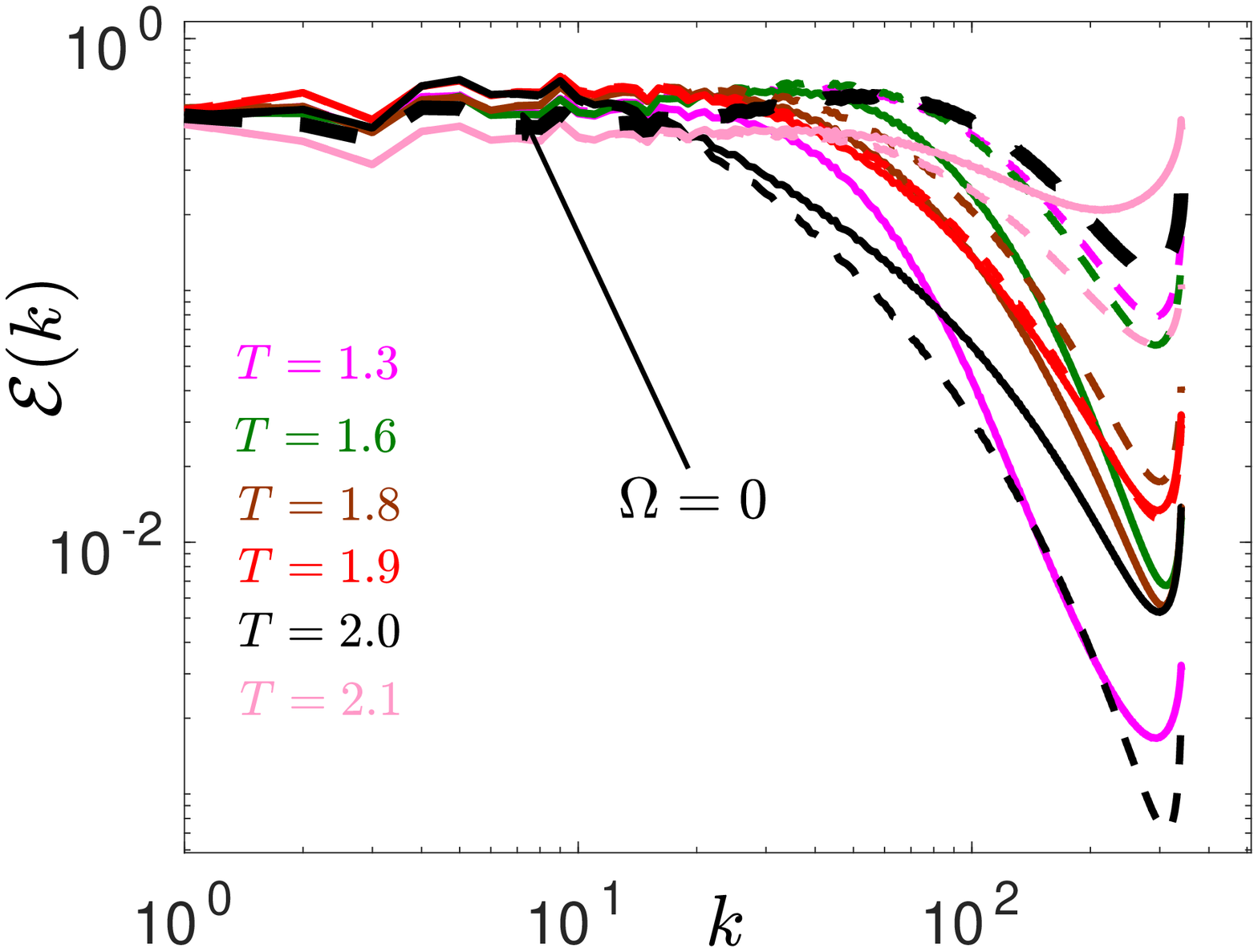}&
   \includegraphics[scale=0.4 ]{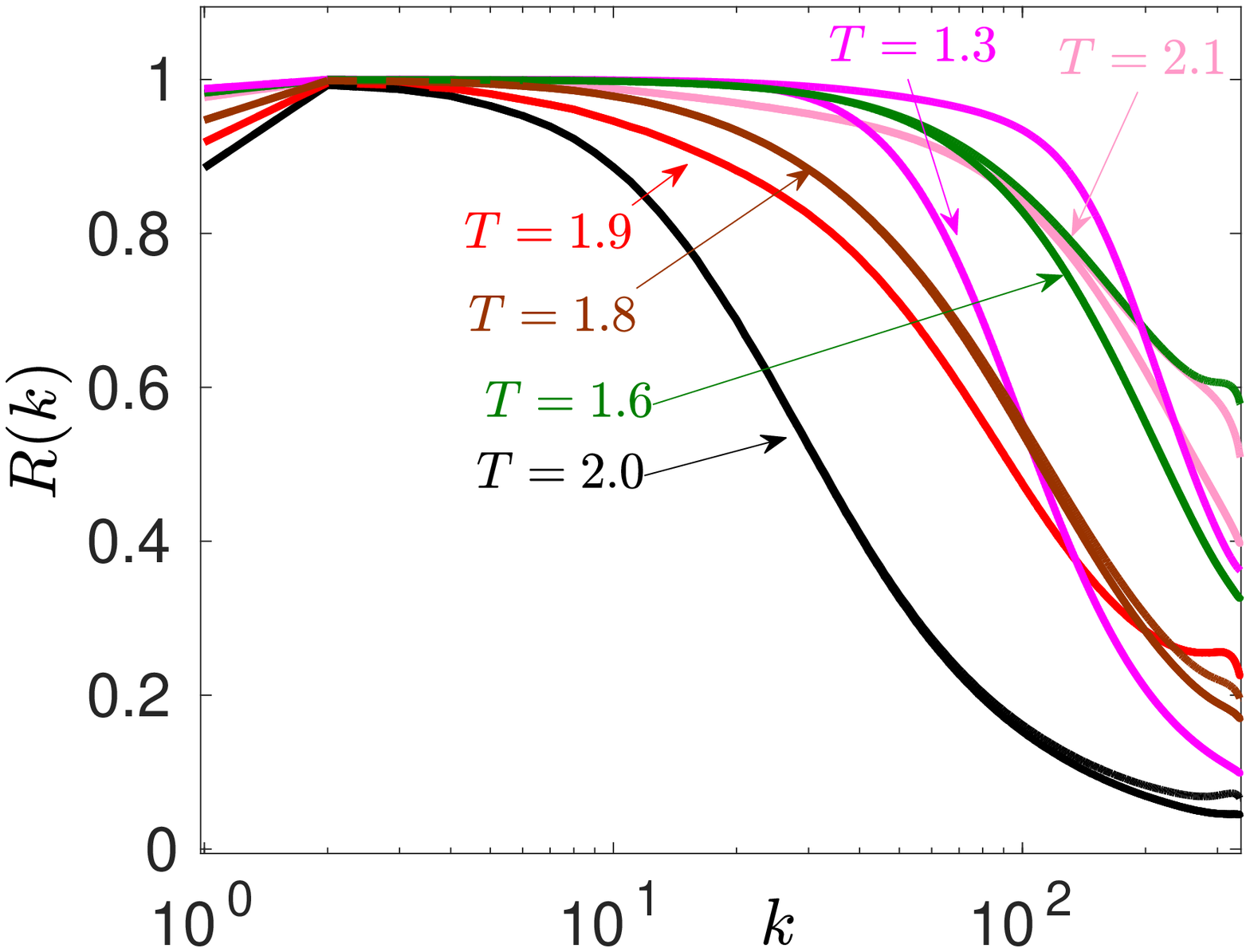} \\
   \end{tabular}
\caption{\label{F:2} Color online. Panel a: Normalized compensated by $k^{5/3}$ energy spectra of normal (solid line) and superfluid (dashed line) components ${\cal E}\sb{n,s}$ [ \Eq{E}]. The  thick black  dashed line, marked $\Omega=0$, corresponds to the energy spectra in the decoupled case. Panel b:   Normalized cross-correlation functions $R_1(k)$ (solid lines) and $R_2(k)$ (dashed lines) [\Eq{R-cross}]. }
\end{figure*}

 \section{\label{s:DNS} Statistics  of $\bm ^4$He turbulence: ~~~    DNS results and their analysis}
\subsection{\label{ss:procedure} Numerical procedure}

We carried out a series of DNSs of coupled HVBK \Eqs{NSE} for normal- and superfluid velocities for different temperatures $T$ using a
fully de-aliased pseudospectral code with resolution $1024^3$ collocation points in a triply
periodic domain of size $L=2\pi$. Table\,\ref{t:1} summarizes the parameters used  in  simulations. The temperature dependencies  of some  parameters used in  \Eq{NSE} [ratio of  superfluid and normal fluid densities $\rho\sb s/ \rho \sb n$,  mutual friction parameters $\alpha$  and $\alpha\rho\sb s/\rho \sb n$  and  the (effective) kinematic viscosities $\nu\sb s$ and $\nu\sb n$ ] are shown in  \Fig{F:1}.

To obtain steady-state evolution,  velocity fields of the normal and superfluid components are stirred by two independent random Gaussian forcings:

\begin{equation}\label{force}\langle {\bm \tilde \varphi}_u({\bm k},t)\cdot { \bm \tilde \varphi}_u^*({\bm q},t') \rangle =\Phi(k) \delta( {\bm k}-{\bm q}) \delta(t-t')
\widehat P({\bm k})\,,
 \end{equation}
 where $\widehat P({\bm k})$ is a projector assuring incompressibility
and $\Phi(k)=\Phi_0 k^{-3}$; $^*$ stands for complex conjugation and the forcing amplitude $\Phi_0$ is nonzero only in a given band of Fourier modes:
 $ k^{\tilde \varphi} \in [0.5,1.5]$. Time integration is performed using  2-nd order Adams-Bashforth scheme with viscous term exactly integrated.

Simulations for the  temperature range $T=1.3-2.1\,$K were carried out with the superfluid viscosity fixed and
the value of $\nu\sb n$ found from the ratio $\nu\sb s/\nu\sb n$  taken equal to the known value of this ratio at each temperature. In addition,  the simulations at low resolution ($256^3$) for $T=1.7-2.1\,$K and at high resolution for high temperature range $T=1.8-2.1$K were carried out also with the constant  viscosity of the normal fluid component  and  $\nu\sb s$  varied in accordance with the temperature dependence of their ratio.
These additional simulations allowed to distinguish the influence of the temperature dependence and the Reynolds number dependence of the structure functions and flatness of two components. When $\nu\sb s$ is constant and $\nu\sb n$ is varied, the results for normal component are affected by both dependencies, while the results for  the superfluid component depend only on the temperature. Situation is reversed when $\nu\sb n$ is fixed and $\nu\sb s$ is varied: in this case the results for the superfluid component depend simultaneously on the changing temperature and Reynolds number.  Our exploratory  results show that the outcome does not depend on the protocol. The detailed behavior needs to be explored further. We comment on this double dependence where relevant.

\subsection{\label{ss:energy}Turbulent energy spectra }

One dimensional energy spectra   for the normal fluid  component (solid lines) and  for the superfluid component (dashed lines) are shown in \Fig{F:2}a.  The spectra are compensated by the Kolmogorov 1941 (K41) scaling behavior and normalized by the total kinetic energy of the normal fluid component:
\begin{equation}\label{E}
\C E\sb {n,s}(k)=k^{5/3}E\sb{n,s}(k)/E\sb n\, ,\quad E\sb n=\int E\sb n(k) dk\ .
\end{equation}
The line colors used for  different temperatures from $T=1.3\,$K to $T=2.1\,$K are the same in all figures.    For comparison we also show in \Fig{F:1}a by thick black dashed line the spectrum corresponding to the classical hydrodynamic turbulence. It was obtained by simulations of the decoupled \Eqs{NSE} with $\Omega=0$ and equal viscosities.

There are several important features of these spectra. First of all, all the compensated spectra (both for normal and superfluid components) have a plateau in the small $k$-range, $k \lesssim 20$ for our resolution, meaning that  $E\sb s(k)\approx E\sb n(k)\propto k^{-5/3}$. We therefore  confirm previous observations\cite{Tabeling,Roche1,Roche2, Rev1,Rev2} that the energy spectra in the coflow of superfluid He are similar to the
energy spectra, observed in classical turbulence\,\cite{Frisch}. We do not resolve the intermittency corrections in the spectra.

For $k \gtrsim 20$, the compensated spectra for different temperatures fall off differently, due to combined influence of the viscous dissipation and dissipation by mutual friction.
The interplay of these two types of dissipation leads to a complicated relation between the normal and superfluid spectra. Since $\nu\sb s < \nu \sb n$ for  $T \lesssim 1.85$K and  $\nu\sb s > \nu \sb n$  for $T \gtrsim 1.85\,$K (see \Fig{F:1}c and the Table \ref{t:1}), the normal spectra (solid lines) decay faster than the superfluid spectra (dashed lines of the same color) for low temperatures and slower for high $T$, almost coinciding for $T=1.9$K (red lines).
In addition, all normal fluid spectra (solid lines in \Fig{F:2}a) lie below the classical spectrum (thick black dashed line for $\Omega=0$) for the same $\nu \sb n$ . The additional energy dissipation is caused by the mutual friction, which becomes important where the velocities of the components become unlocked.
On a semi-quantitative level this behavior is similar to that found earlier in \Ref{PRB} using Sabra-shell model approximation.

\begin{figure*}
 \begin{tabular}{ccc}
  (a) T=1.3 &  (b) T=1.8 &  (c) T=2.1 \\
\includegraphics[scale=0.27 ]{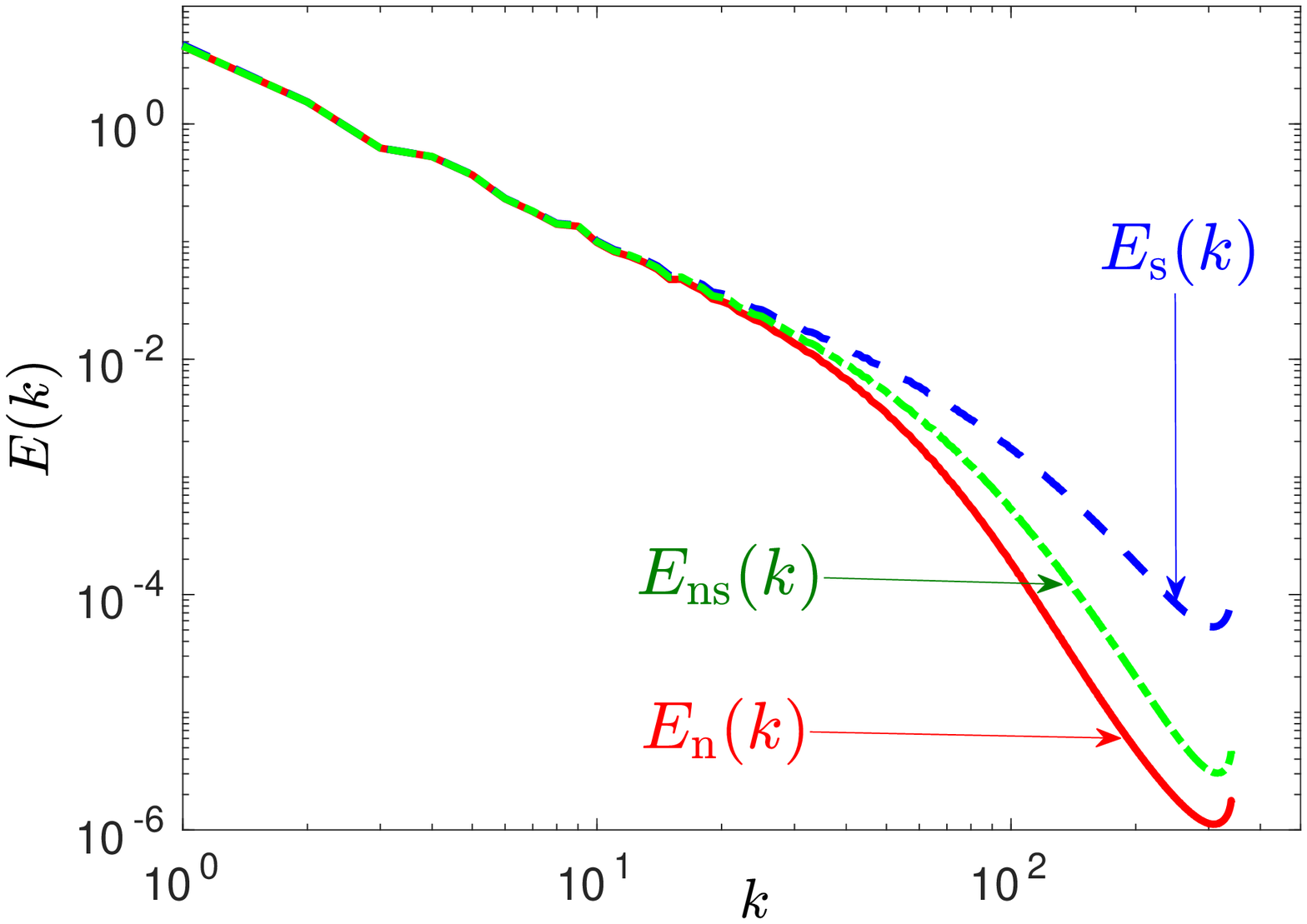}&
\includegraphics[scale=0.27 ]{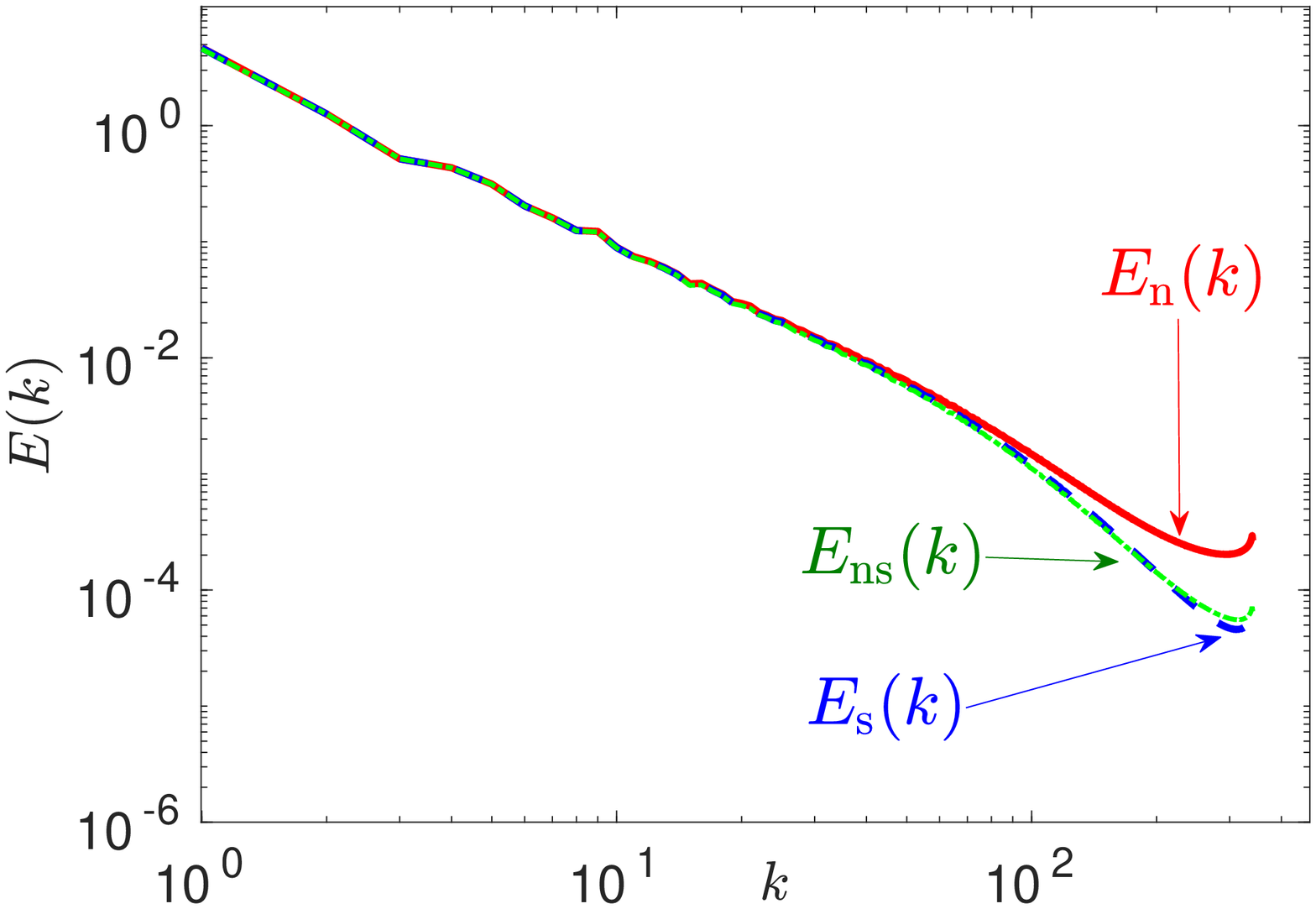}&
   \includegraphics[scale=0.27 ]{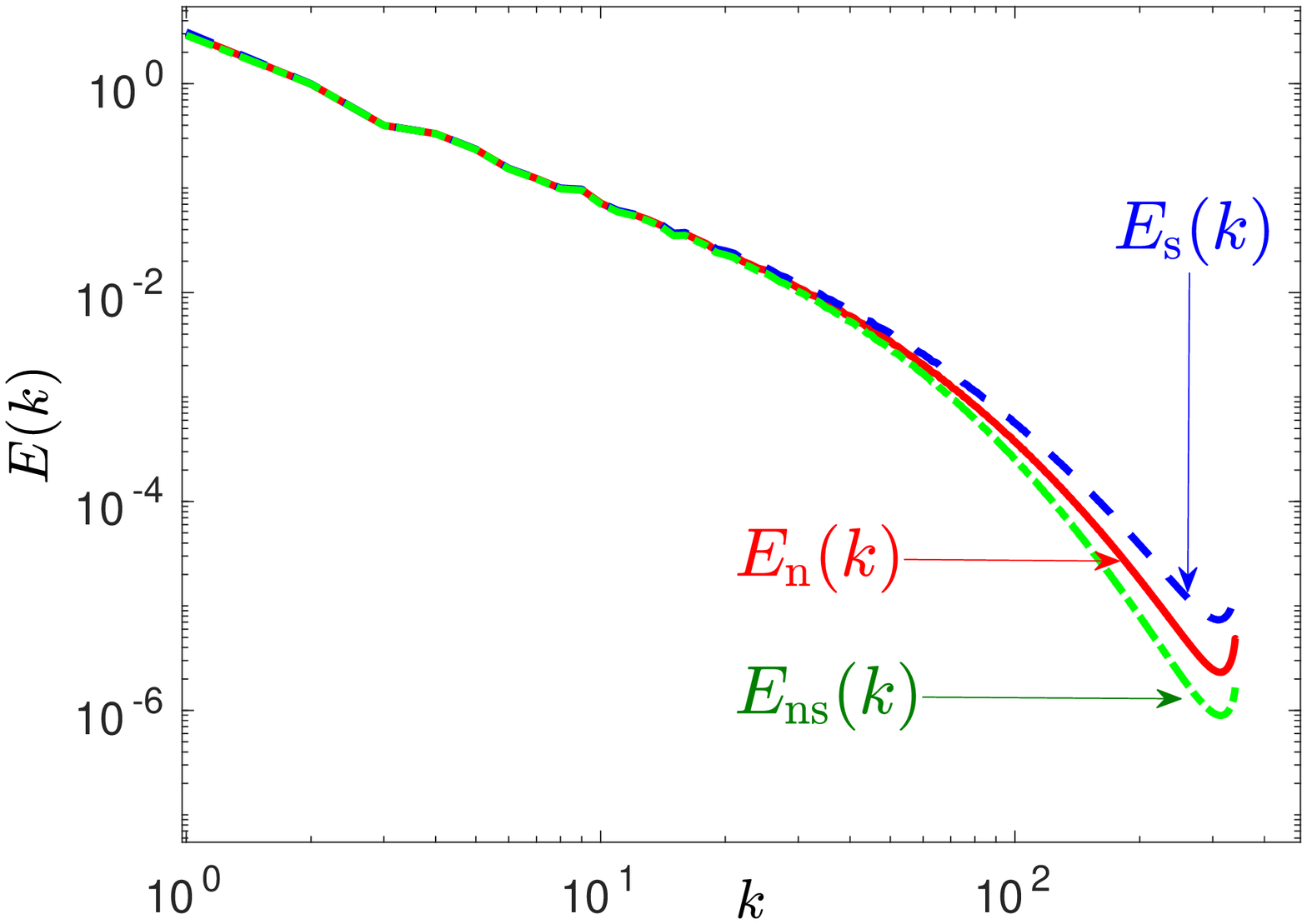}\\
 \end{tabular}
 \begin{tabular}{ccc}
(d) T=1.3 &  (e) T=1.8 &  (f) T=2.1 \\
  \includegraphics[scale=0.27 ]{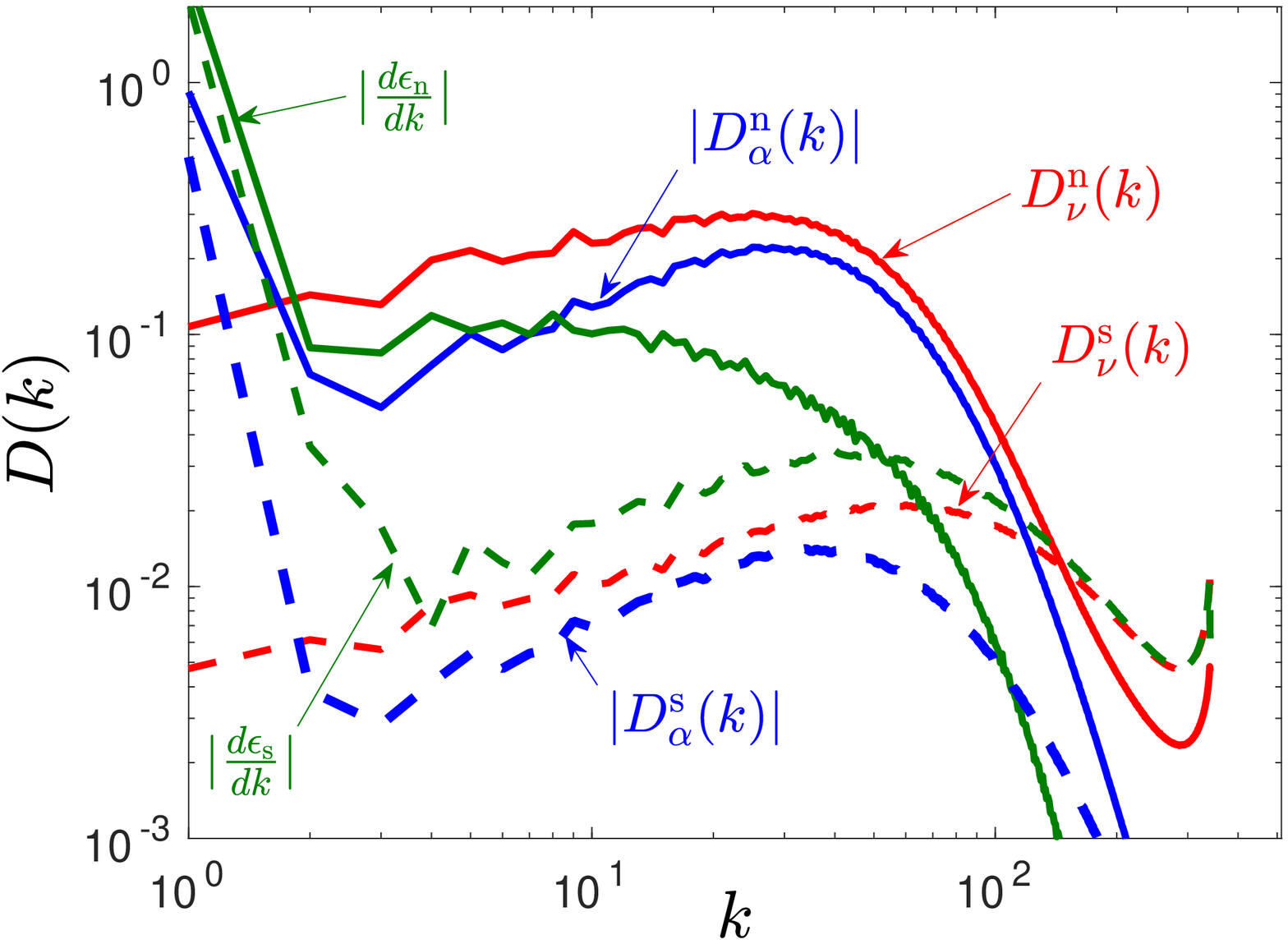}&
 \includegraphics[scale=0.27 ]{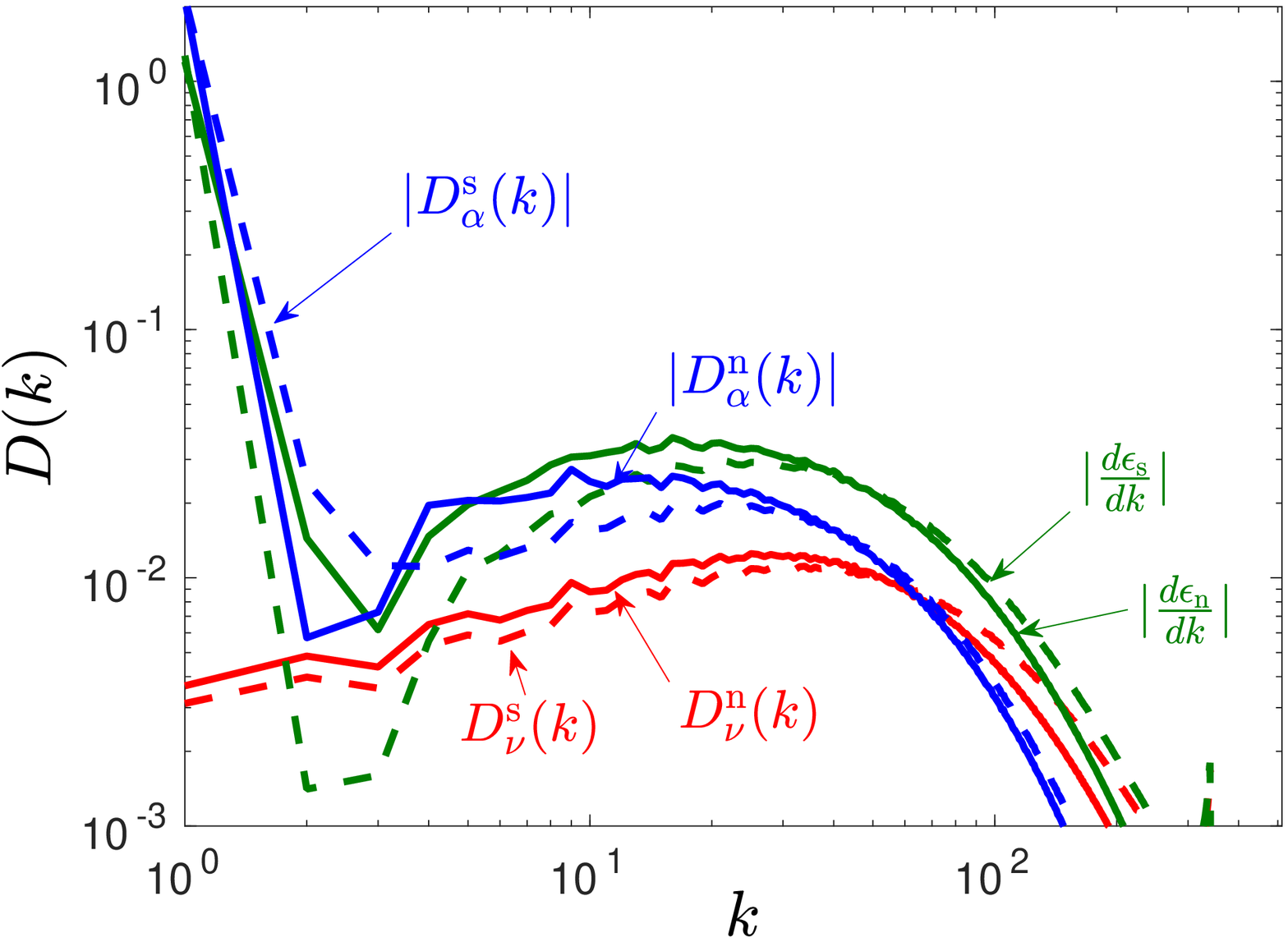}&
  \includegraphics[scale=0.27 ]{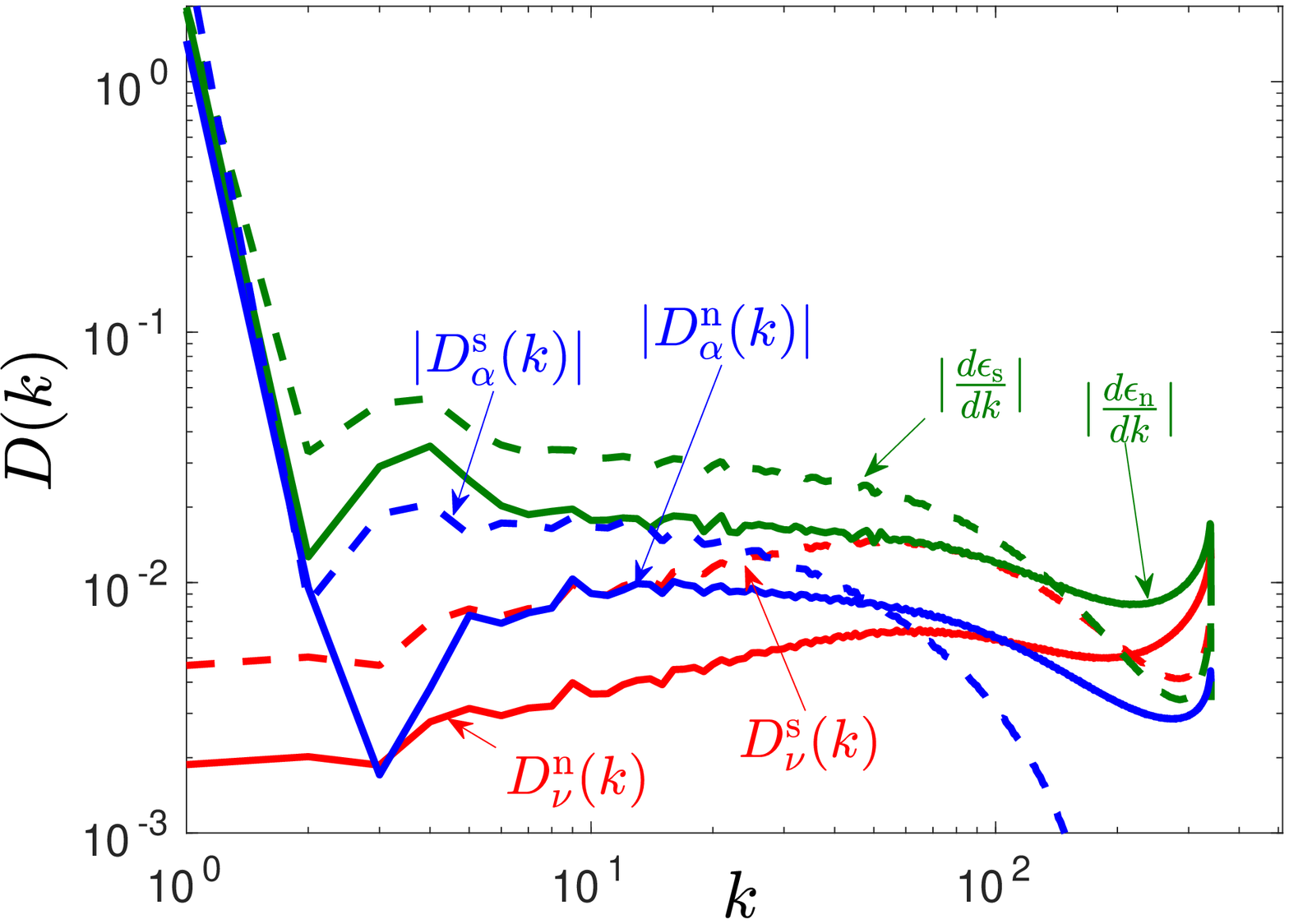}\\
   \end{tabular}
    \caption{\label{F:3}Color online.  Energy spectra [Panels (a)-(c)] and  the energy balance [Panels (d)-(f)]  for the normal (solid lines) and superfluid (dashed lines) components.
    The spectra and balances are shown for three representative temperatures. Different terms of \Eq{BAL} are marked in the panels.  In panels (d)-(f), absolute values of the terms are shown.
    }
\end{figure*}

\subsection{\label{ss:cross} Cross-correlation   of the normal and superfluid velocities}
To better understand the behavior of the energy spectra  at different temperatures, we consider  correlation between normal and superfluid velocities.
 It is often  assumed \cite{VN}  that the normal and superfluid velocities are ``locked" in the sense that

 \begin{equation}\label{loc}
 \B u\sb n (\B r,t)= \B u\sb s (\B r,t)\ .
 \end{equation}

To quantify  the statistical grounds for this assumption,  we use  the 1D cross-velocity correlation function
$\C E\sb{ns}\propto \< \B u\sb n (\B k)\cdot \B u\sb s (\B k)\> $, \Eqs{Xcorr},  normalized in two ways \cite{LNS-2006,PRB,LP-QFS}:
\begin{subequations}\label{R-cross}
\begin{eqnarray}\label{crossA}
R_1(k)&=&\frac {2\, \mbox {Re}\{ E\sb{ns}(k)\}}{E\sb s (k)+E\sb n (k) }\,,\\
\label{crossB}
R_2(k)&=&\frac { \mbox {Re}\{ E\sb{ns}(k)\}}{\sqrt{ E\sb s (k)\cdot E\sb n (k)} }\ .
\end{eqnarray}\end{subequations}
 Here Re\{...\} denotes real part of a complex variable.

 Both cross-correlations are equal to unity for fully locked superfluid and normal velocities [in the sence of  \Eq{loc}],   and both vanish if the
 velocities are statistically independent. However, if the velocities are proportional to each other
  \begin{subequations}\label{slave}
\begin{equation}\label{slaveA}
 \~{\B u}\sb  n (\B k,t)=C(k)\~{\B u}\sb s (\B k,t)\,,
  \end{equation}
 with $C(k) \ne  1$,
then $R_1 (k)= 2C(k) / (C^2(k)+1)<1$, while $R_2 (k)$ is still equals to unity, $R_2 (k)= 1$. In any case $R_1 (k) \leqslant R_2 (k)$.

The cross-correlations  $R_1 (k)$ and $R_2 (k)$ are shown in \Fig{F:2}b by solid and dashed lines, respectively,  with the same color code for different $T$ as in  \Fig{F:2}a.  Clearly, both $R_1 (k)$ and  $R_2 (k)$ monotonically decrease with $k$ and for  some cases become significantly smaller than unity already at $k\approx 10$ . For example,  for $T=2.0\,$K and $k>50$,  $R_1 (k)\approx R_2 (k)<0.3$. Thus the normal and superfluid velocities begin to decorrelate in the cross-over region between inertial and viscous interval and are  practically uncorrelated  in the viscous subrange. For temperatures around $1.8\div 2.0\,$K, the  normal- and superfluid viscosities are similar  and  $R_1 (k)\approx R_2 (k)$ at all $k$.

On the other hand,  for low and high $T$,   $R_2 (k)$  significantly exceeds  $R_1 (k)$, especially in the large wavenumber limit. This is best visible for $T=1.3\,$K. At this temperature
$\nu\sb n \simeq 25 \nu\sb s$ and the normal fluid component is  overdamped at large $k$: $v\sb n(k)\ll v\sb s(k)$.  As a result, $v\sb n(k)$ does not have its own nonlinear dynamics.  Accounting in \Eq{NSEn} (in  $\B k$-representation) only for the viscous and mutual friction terms  we get:
\begin{equation}\label{slaveB}
  \nu\sb n k^2 \~ {\B u}\sb n(\B k,t)\approx \frac {\alpha \rho \sb s\Omega }{\rho \sb n}\~ {\B u}\sb s(\B k,t)\ ,
\end{equation}
 meaning that the normal fluid velocity  follows the superfluid one in the sense of
\Eq{slaveA}
with
\begin{equation}\label{slaveC}
C(k)= \frac {\alpha \rho \sb s\Omega }{\rho \sb n\nu \sb n k^2}<1 \ .
\end{equation}
  \end{subequations}
  In this approximation $E\sb n (k)\approx [C(k)]^2 E\sb s (k) <  E\sb s (k)$ and $R_1(k)<1$, while $R_2(k)\approx 1$. Similar (but less pronounced)  effect takes place at temperatures near the $T_\lambda$, where $\nu\sb s \gg \nu\sb n$, see for example $R_1(k)$ and $R_2(k)$ in \Fig{F:2}b for   $T=2.1\,$K, for which $\nu\sb s \simeq 4 \nu\sb s$. The fast change  in the component's viscosity and density for $T>2\,$K [see \Fig{F:1}(b) and (c)] leads to striking difference in all statistical properties of superfluid $^4$He at $T=2\,$K and $T=2.1\,$K, shown in the figures by black and pink lines, respectively.

We stress that numerical results shown in \Fig{F:2}b qualitatively agree for  most of temperatures with the analytical  expression of the cross-correlation $ E\sb {ns}(k)$\,\cite{LNS-2006,PRB}, which in current notations reads:
\BE{lim1}    \C E\sb{ns}(k) = \frac {\a\,  \O    [\r \sb n \,    E\sb n(k)+ \r \sb s \,  E\sb s(k) ] }
{\a \,\O  \r+ \r\sb n [(\nu\sb s+ \nu\sb n)\, k^2 + \g\sb n(k)+ \g \sb s(k)]   }\ . \ee
Here  $\g \sb s(k)=  \sqrt{k^3  E\sb s(k)} $ and $\g \sb n(k)=  \sqrt{k^3  E\sb n(k)} $  are dimensional K-41 estimates of the  turnover frequencies of eddies in the superfluid and normal fluid components, respectively.

\begin{figure*}
\begin{tabular}{cc}
 \large Normal fluid & \large Superfluid \\
 (a) & (b) \\
   \includegraphics[scale=0.4 ]{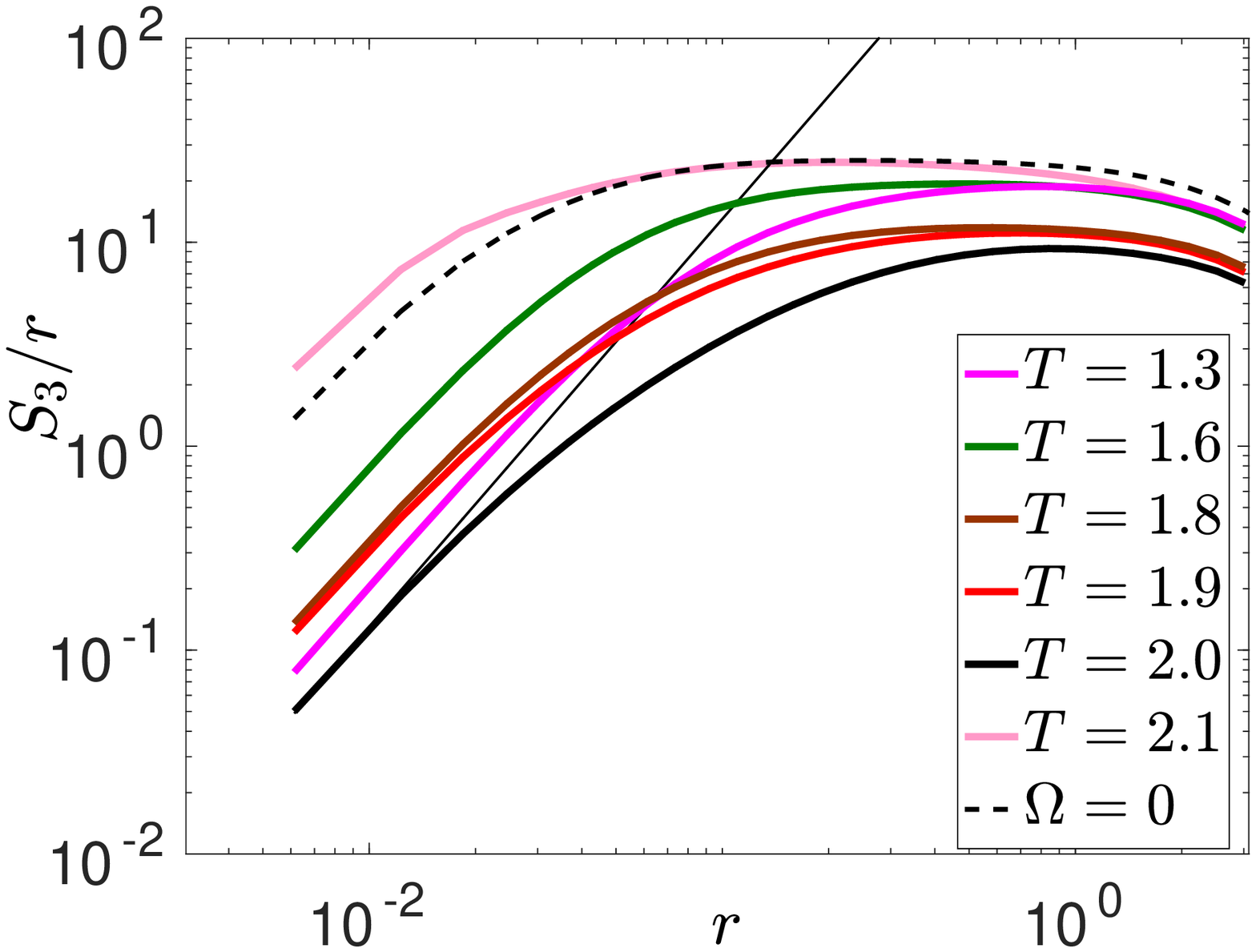}&
  \includegraphics[scale=0.4 ]{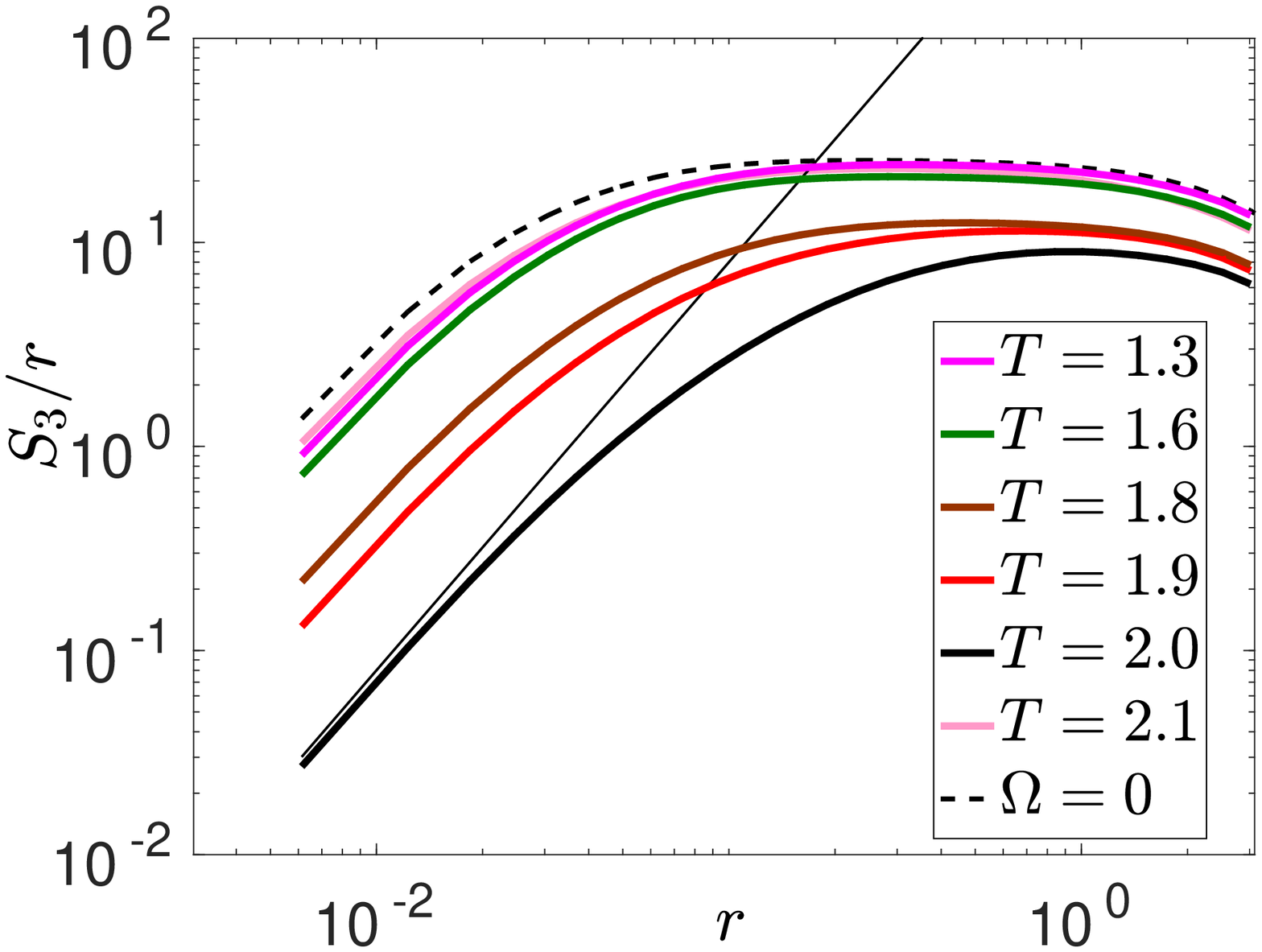}\\
   \end{tabular}
  \caption{\label{F:4}  Color online.  The compensated structure functions $S_3(r)/r$ of normal [Panel (a)] and superfluid [Panel (b)] components as a function of $r$. Here and in other figures, the thick black dashed line, marked $\Omega=0$, corresponds to the decoupled case.  The thin straight solid lines indicate  viscous behavior $S_3\propto r^3$}.
\end{figure*}

\begin{figure*}
 \begin{tabular}{cc}
  \large Normal fluid & \large Superfluid \\
 (a) & (b) \\
     \includegraphics[scale=0.4]{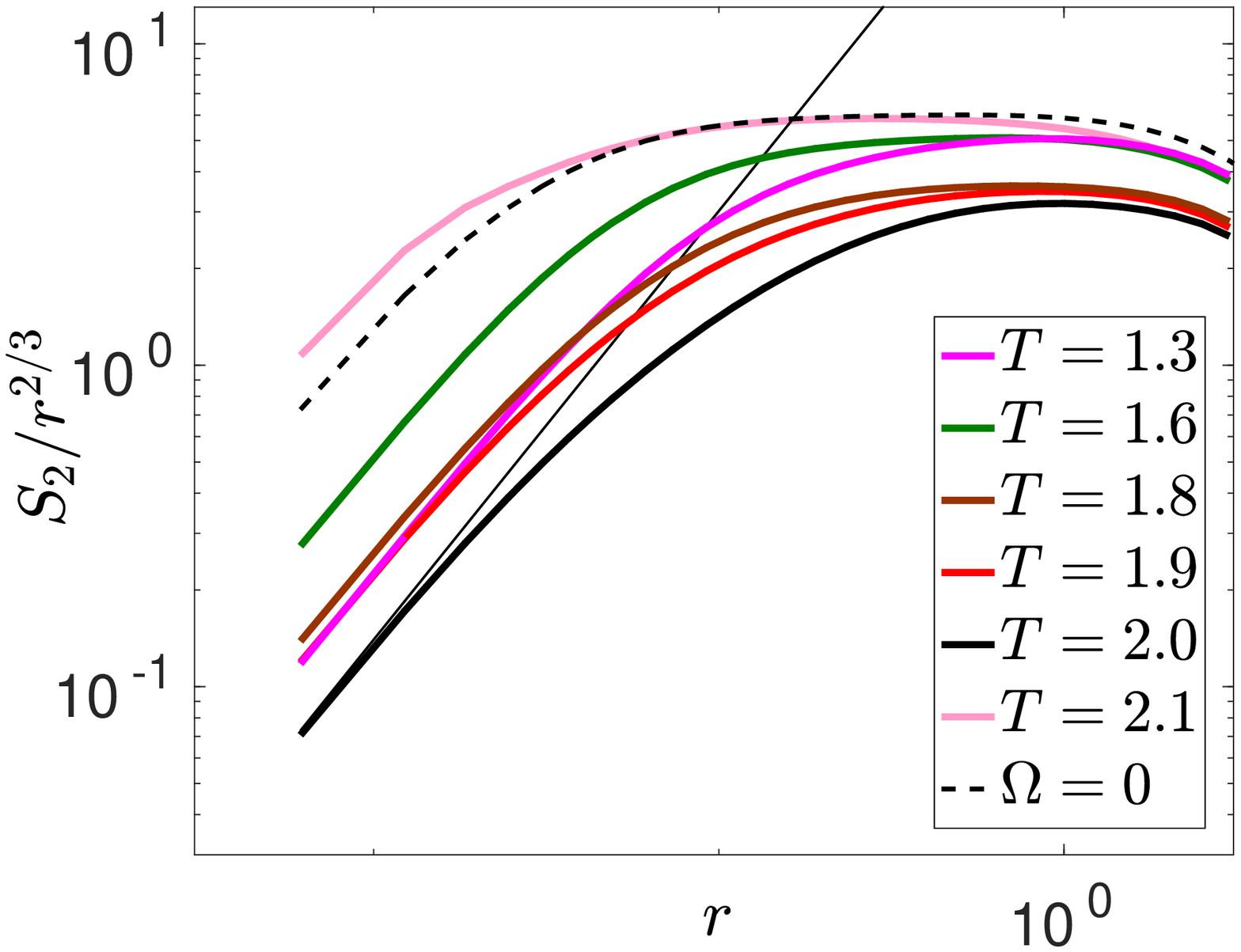}&
   \includegraphics[scale=0.4 ]{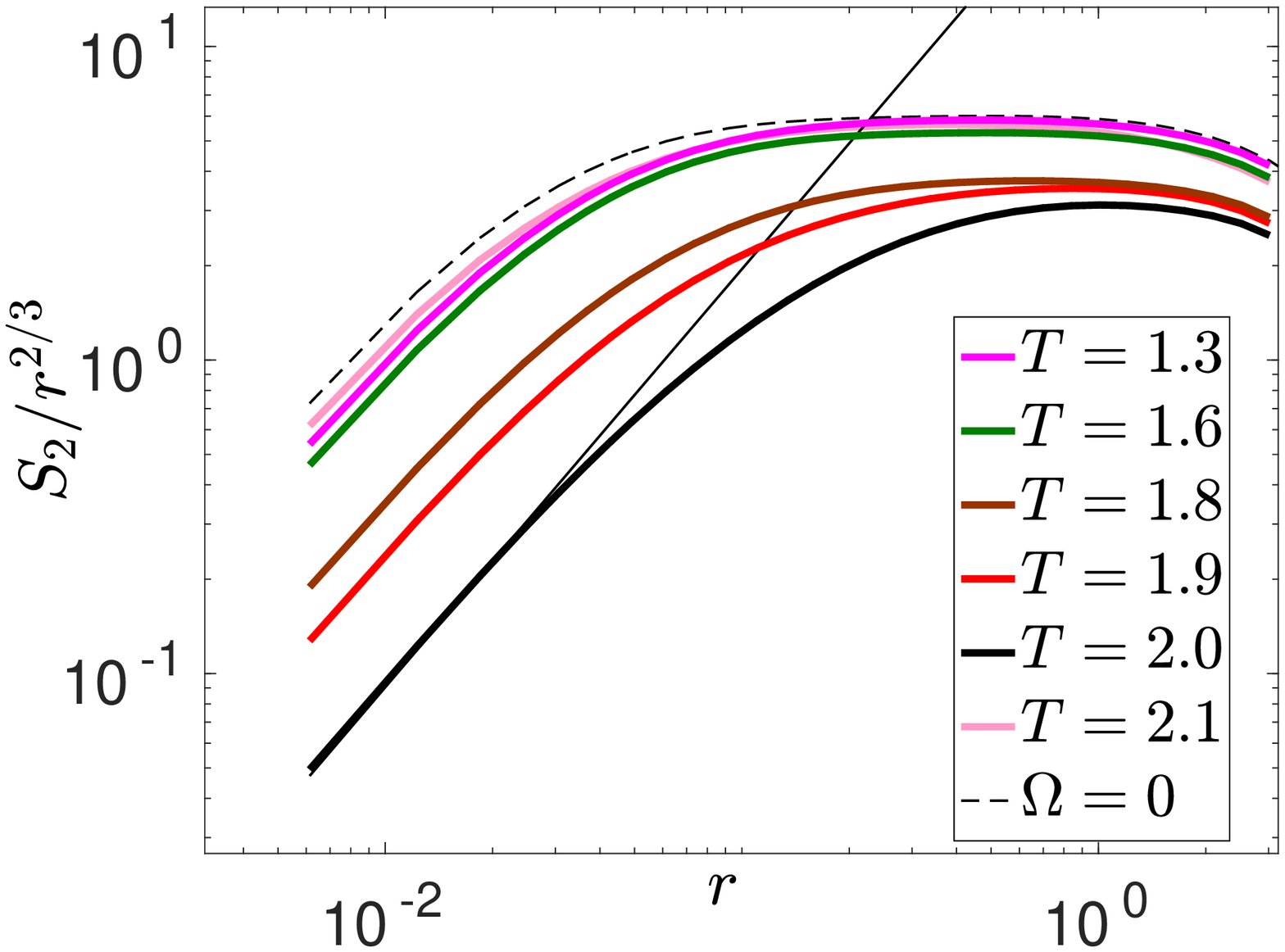}  \\
 (c) & (d)  \\
   \includegraphics[scale=0.4 ]{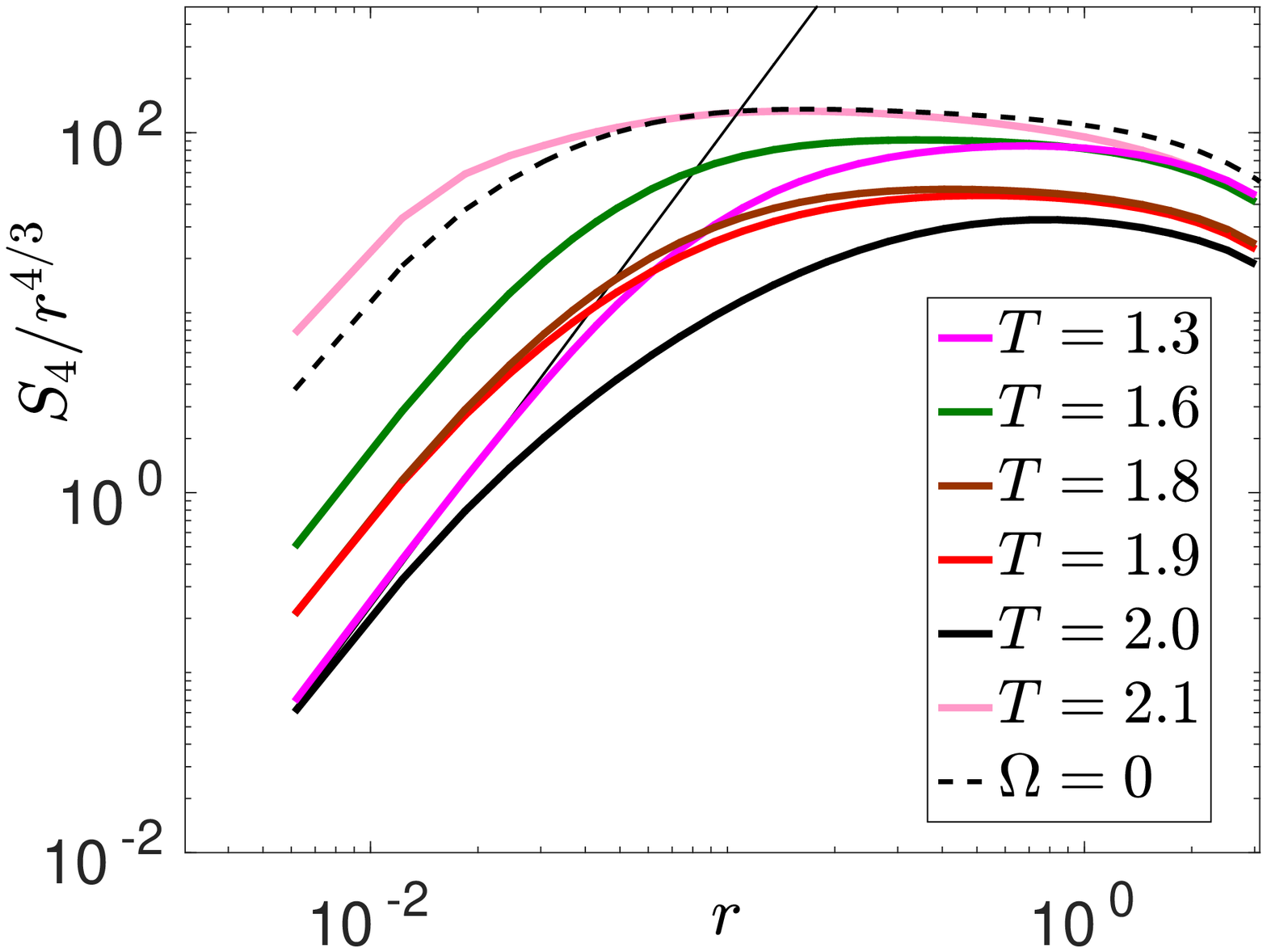}&
   \includegraphics[scale=0.4 ]{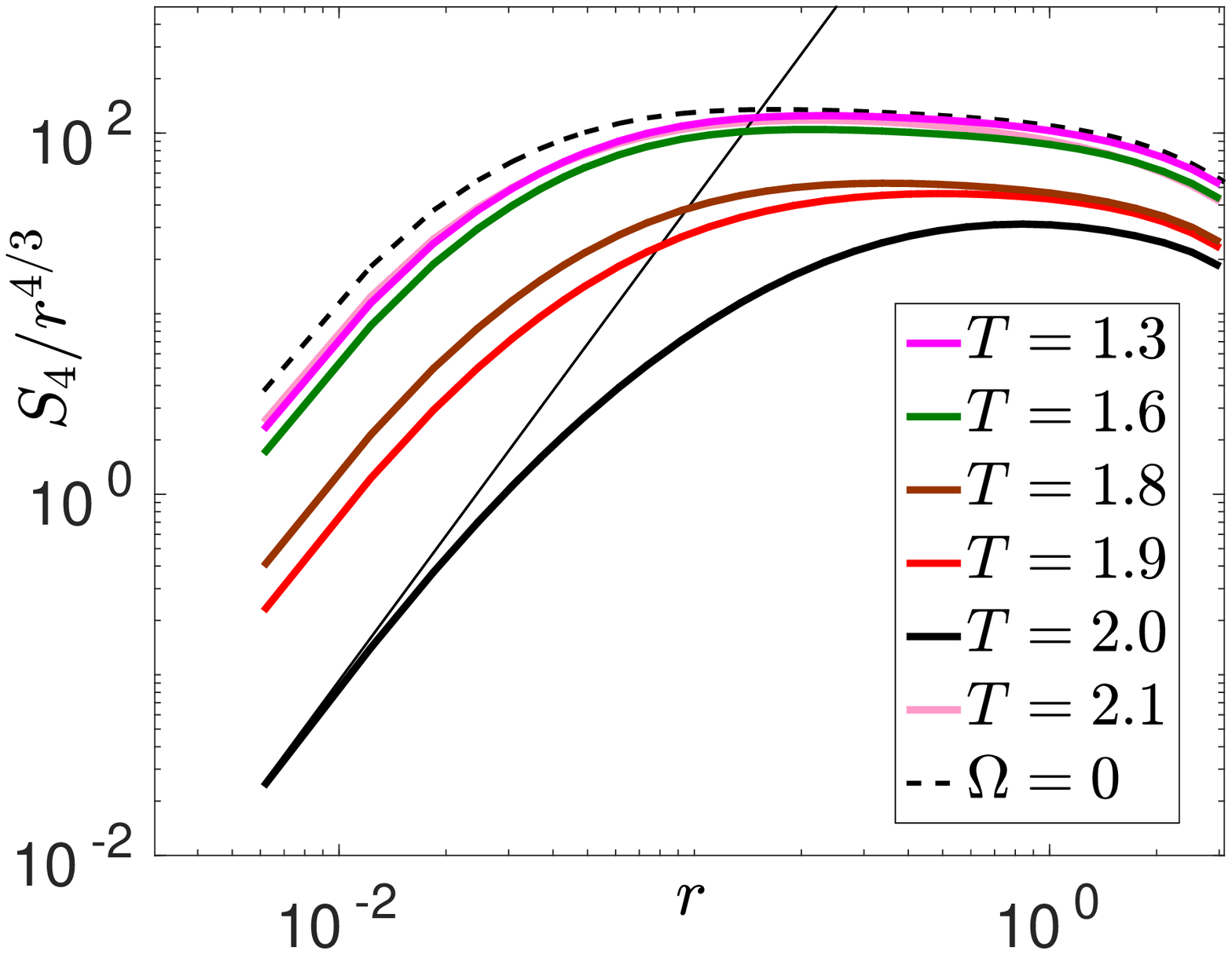}  \\
 \end{tabular}
  \caption{\label{F:5} Color online. The compensated structure functions $S_2(r)/r^{2/3}$ [Panels (a),(b)]  and  $S_4(r)/r^{4/3}$ [Panels (c),(d)]  as a function of $r$. The thin straight solid lines indicate viscous behavior $S_2\propto r^{2}$ and $S_4\propto r^4$.
     }
\end{figure*}

\begin{figure*}
\begin{tabular}{cc}
\large Normal fluid & \large Superfluid \\
 (a) & (b)  \\
   \includegraphics[scale=0.45 ]{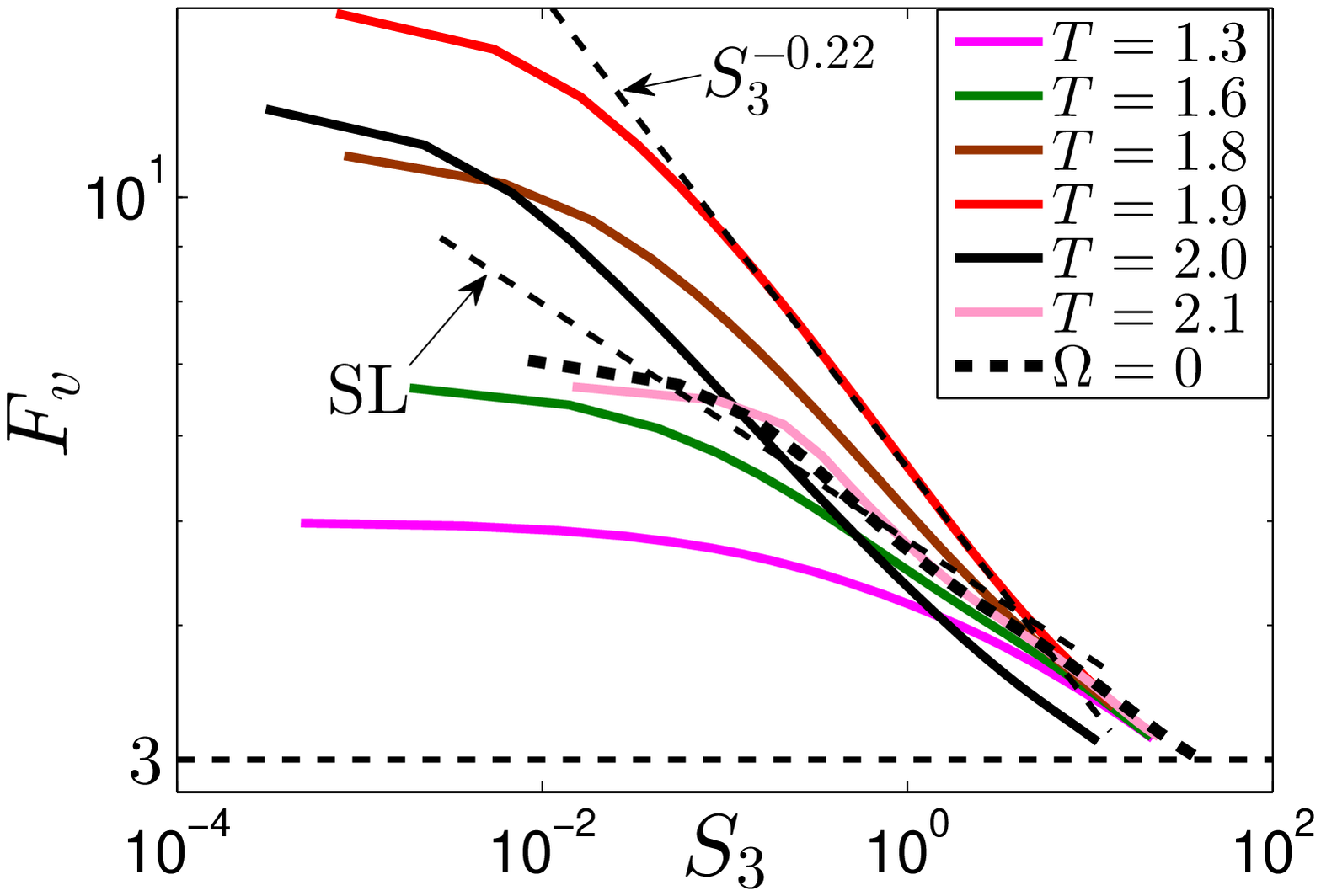}  &
   \includegraphics[scale=0.45]{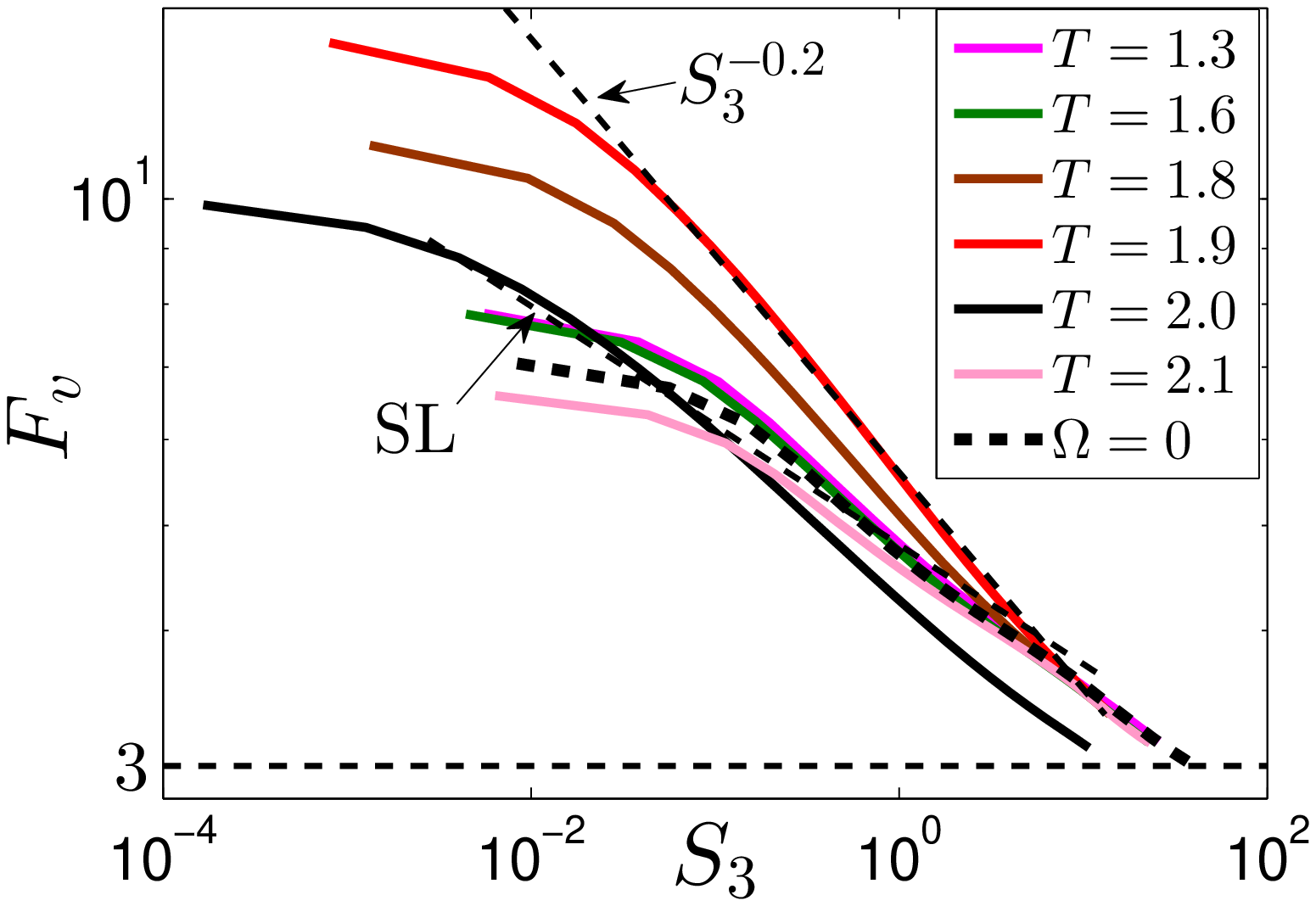}   \\
    \end{tabular}

  \caption{\label{F:7} Color online.  Flatness of normal [Panel(a)] and  superfluid [Panel(b)] components as a function of $S_3$. The horizontal dashed line correspond to $F\sb v=3$.  Note log-log scale.  Thin straight dashed lines indicate the scaling, corresponding to the She-Leveque model of classical turbulence \Ref{SL}(marked as "SL") and the fit for structure function at $T=1.9$K.}
\end{figure*}

\begin{figure*}

 \begin{tabular}{cc}
 \large Normal fluid & \large Superfluid \\
  (a)& (b)  \\
   \includegraphics[scale=0.5 ]{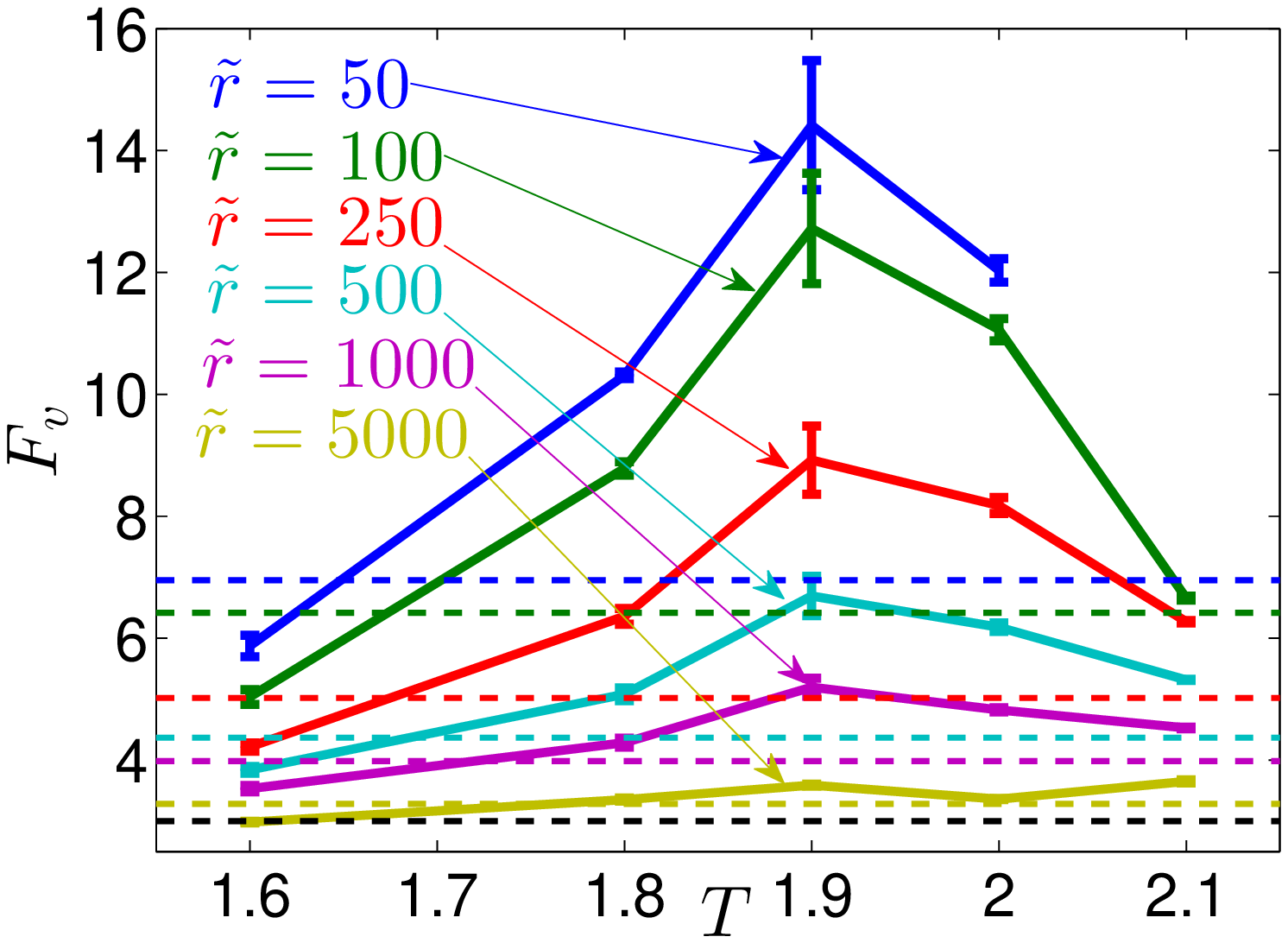}&
  \includegraphics[scale=0.5 ]{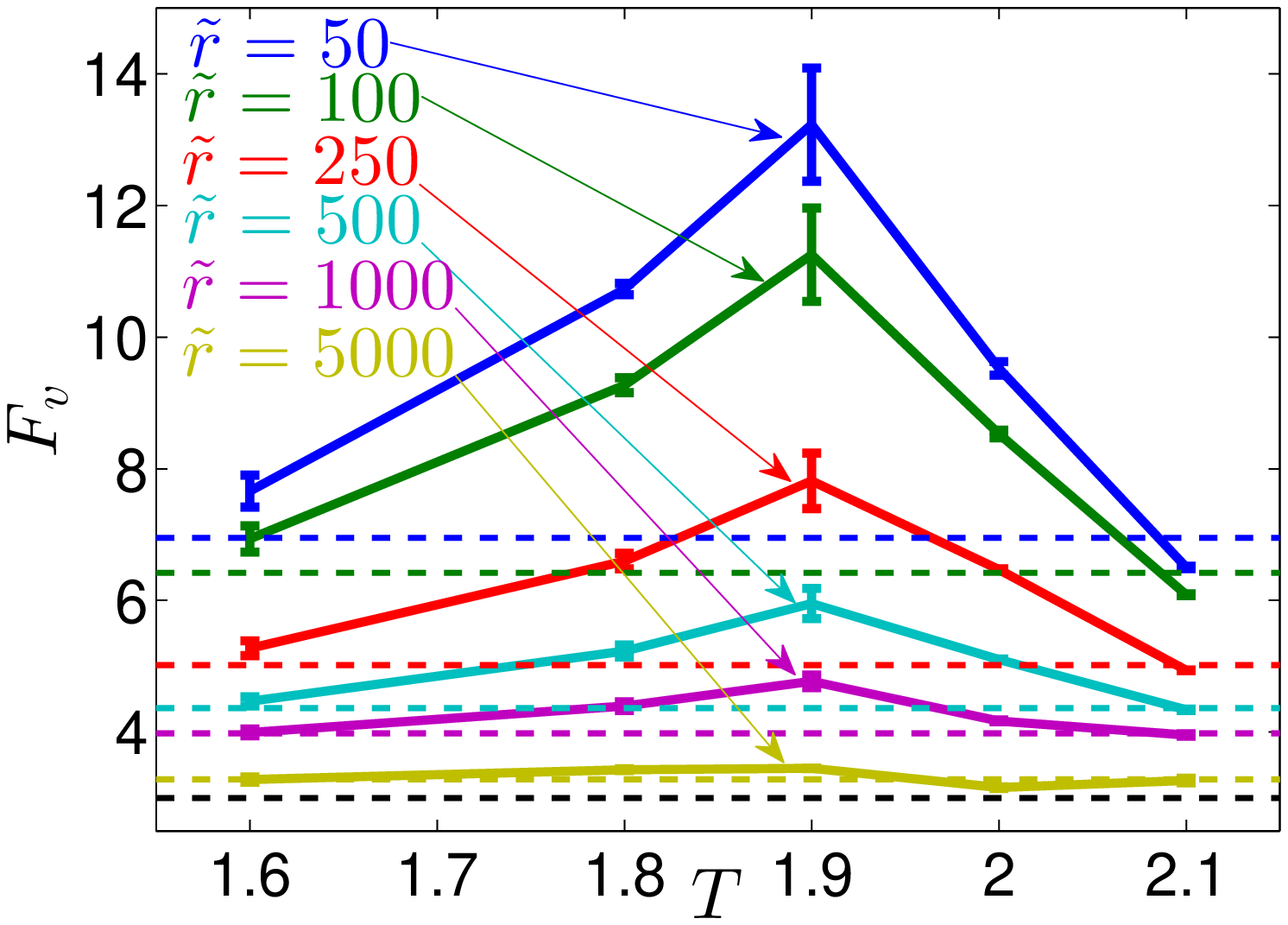} \\
\end{tabular}
\caption{\label{F:8} Flatness as a function of temperature for the normal component [Panel (a)] and  for the superfluid component [Panel (b)] for different separations $\tilde{r}\sb{n,s}=r/\eta\sb{n,s}$. The horizontal dashed lines lines correspond to the classical values of flatness for the same separations (marked by the same colors). The classical values are represented by the  decoupled case. The lowest black dashed line corresponds to the Gaussian value $F\sb v=3$.
The erorbars were obtained by averaging results for different intervals of the time realization. }
\end{figure*}


\subsection{\label{ss:eb} Energy balance}
{To quantify the relative importance of different terms of the energy balance equations \Eq{BAL}, we plot them in \Fig{F:3} for three typical temperatures $T=1.3, 1.8$ and $2.1\,$K together  with the energy spectra  $E\sb s,E\sb n$  and the cross-correlation $E\sb{ns}$. {Here the terms  $\langle \Omega(t)E(k,t)\rangle$ in \Eqs{BAL} were calculated directly, not using product of averages.

Starting with low temperature [$T=1.3\,$K, \Fig{F:3}(a),(d)], we note that the normal and superfluid energy spectra and the cross-correlation almost coincide for $k\lesssim 20$. For larger wavenumbers $E\sb n< E\sb{ns}$ and $D_{\alpha,\rm n}$ is negative, transferring energy from the  superfluid component to the normal component. Most of this energy is dissipated by the viscous friction $D_{\nu,\rm n}$ even at low $k$. The normal component is mostly responsible for the energy  dissipation at low $T$ due to large ratio of densities and viscosity.

At moderate temperatures [$T=1.8\,$K, \Figs{F:3}(b) and (e)]
the two components have similar behavior. The energy exchange and dissipation due to mutual friction in both components is dominant over almost all scales, while viscous dissipation takes over only deep in the viscous  $k$-range.

At  high $T$ [\Fig{F:3}(c,f)], the superfluid components is the most dissipative part of the system, by the mutual friction in the inertial range and by viscosity at higher $k$.

\subsection{\label{ss:str} Velocity structure functions}
Statistical properties of the turbulent flows are usually characterized by velocity structure functions $S_n(\bm r)$. Their scaling behavior in the classical turbulence is well known. Much less is known about structure functions in the superfluid He. In this Section we discuss the  velocity structure functions of normal and superfluid components as well as the flatness to analyse  intermittency effects in the two-fluid system.

\subsubsection{Third-order velocity structure functions $S_3$}

One of the most solid results in  hydrodynamics turbulence is the scaling of the third order velocity structure function $S_3(\bm r)\propto \bm r$, which is a consequence of the $\displaystyle \frac45$th law. To see, in which sense this result is valid in the superfluid  He, we  plot  in \Figs{F:4} the normalized $S_3(r)/r$ vs $r$  for both components.  First of all, we notice that  the range of wavenumbers,  in which the expected scaling ($\propto r$ ) for classical(decoupled) case  is observed, corresponds well to the scaling interval of the energy spectrum for this case. Very similar behavior is observed in $S_3$ of both components for the highest available temperature $T=2.1\,$K. At two lowest temperatures, $T=1.3$ and $1.6\,$K, the superfluid structure function $S\sb{3,s}$ almost coincides with the results for $T=2.1$K, while for the normal component $S\sb{3,n}$ is closer to the moderate temperature ones. This is the consequence of the combined influence of the temperature and Re-number dependence of the flow: the results presented here were obtained with fixed $\nu_s$ and for the superfluid component are affected only by temperature. The results for normal fluid obtained with fixed $\nu\sb n$ are very similar to shown here for the superfluid components, with the  difference that the structure function for $T=2.1\,$K does not approach $S_3$ for the decoupled case.

For the moderate temperatures $T=1.8-2.0\,$K, the extent of the inertial interval where $S_3/r$ is horizontal, is shortened.  For $T=2.0\,$K, there is no  inertial scaling range at all, unlike the energy spectrum, in which the inertial range is clearly observed for $k\lesssim 20$.  On  the other hand, a viscous  scaling range with  scaling $\propto r^3$ at small $r$ is observed in this temperature interval.

\subsubsection{Velocity structure functions $S_2$ and $S_4$}

Next we turn to the second and forth order structure functions for both components. In \Fig{F:5} we plot $S_2$  and $S_4$  compensated by their  respective Kolmogorov scaling. Clearly, all main features are similar to those of $S_3$ - for low $T$ and $T=2.1\,$K the structure functions  of the superfluid component are similar to the classical case, while in the intermediate temperature range the inertial range scaling is lost with a scaling range for small $r$ seen for $T=2.0\,$K. Here this  scaling is  $ S_2(r)\propto r^{2}$ and  $S_4(r)\propto r^{4}$ for both components and corresponds to the viscous range scaling. The signature of the mixed temperature and Re-number dependence in the normal fluid component is also very similar to that of the third order structure function. We therefore can hope that using Extended Self-Similarity\cite{ess}(ESS) i.e. using $S_3$ instead of $r$, will extend the scaling range and allow us to have better understanding of the intermittency corrections in superfluid turbulence.

\subsubsection{\label{ss:fl} Flatness and Intermittency }

The most direct information on the intermittency   may be obtained from flatness $F\sb v\= S_4/S_2^2$. In \Fig{F:7} we show $F\sb v$ for the normal and superfluid components  as a function of $S_3$.  Evidently, the behavior of the flatness for the decoupled case resembles that of the classical turbulence: $F\sb v\approx 3$ at large scales and is larger for small scales, reaching the value of about 7 for $r\approx 10^{-2}$. Again, the flatness for both superfluid and normal components at low $T$ and for $T=2.1\,$K is close to the decoupled case, while in the intermediate temperature regime the flatness grows faster towards small scales (with an apparent exponent $-0.2$ for $T=1.9\,$K compared to $-0.14$ for the decoupled case) and reaches values above 10. These observations confirm that intermittency in the intermediate range of temperatures is stronger that in classical turbulence.

Since accurate extraction of scaling exponents is difficult at our resolution, we plot in \Fig{F:8} the values of flatness at a number of normalized scales $\tilde r\sb{n,s}=r/\eta\sb{n,s}$, where $\eta\sb{n,s}=\sqrt{2}\nu\sb{n,s}/u\sb{rms}$, for different temperatures. The erorrbars were calculated by averaging the values of flatness obtained over different parts of the time realization. The horizontal dashed lines mark the values of flatness in the classical turbulence ( represented here by the decoupled case). The color code of the lines indicated the scale for which $F\sb v$  was calculated. Clearly, the deviation from  Gaussianity is close to that in the classical hydrodynamic turbulence at large scales and stronger for small scales. This intermittency enhancement is particulary notable  for $T=1.8-2.0\,$K, with small scales flatness exceeding twice the classical values for both components.
The smaller-than-classical values of $F\sb v$ for normal component at $T=1.6\,$K re due to mixed influence of temperature and Re number dependence of the structure functions. These values are similar to $F\sb v$ for superfluid component when calculated with fixed $\nu\sb n$, although the errorbars in this case are larger.

\subsection{\label{ss:i-enh}Flip-flop scenario of the intermittency enhancement in \He4 turbulence}
The intermittency enhancement, clearly demonstrated in \Fig{F:8}, takes place at temperatures, for which  properties of normal and superfluid components are very similar. Closeness of densities leads to most efficient energy exchange by mutual friction. Indeed, the dissipation by mutual friction is almost identical at $T=1.8\,$K (cf.\Fig{F:3}) and is responsible for the energy dissipation in both components at almost all scales.
We therefore suggest a variant of a \emph{flip-flop scenario}\cite{DNS-He3} of the intermittency enhancement in \He4 turbulence by a  random energy transfer between normal- and superfluid components due to mutual friction. Such an energy exchange serves as additional random forcing in a wide range of scales, leading to enhanced intermittency of the velocity fluctuations.

\section{\label{s:sum} Summary}
We performed a series of DNS of the two-fluid gradually damped HVBK equations for homogeneous isotropic coflows in superfluid  $^4$He. The two fluid components are interacting via a self-consistent and non-linear mutual friction, defined in terms of a temperature dependent coupling factor which is a function of the superfluid enstrophy spectrum. The statistical properties of both components, characterized by  energy spectra, velocity  structure functions and flatness, are similar, although not identical. We found that the two components are less correlated in the range of temperatures where their densities and viscosities are close. On the other hand they are  more correlated when the density of one of the components dominates. One can understand this as ``slaving" of the rare component by the dense one that dominates the composition. When the two components are close in densities each can have
a life of its own, reducing the measured correlations between them.

A significant enhancement of small scale intermittency, characterized by flatness, is observed in the intermediate temperature range $T=1.8-2.0\,$K. We suggest a flip-flop mechanism of such an enhancement. The efficient energy exchange between two components by mutual friction serves as an effective forcing on a wide range of scales. This forcing effectively intervenes with the energy cascade over scales. This effect is simultaneously  present in both components. This observation confirms previous numerical results \cite{He4,GPEgrid}.

Given  the present available resolution, it is difficult to make any systematic assessment about existence of pure inertial range scaling exponents which is independent of the Reyonlds number. Accordingly we cannot state whether these are different from the ones measured in homogeneous and isotropic classical turbulence. Usual phenomenology would predict that the mutual friction should induce some sub-leading scaling corrections that might indeed be the reasons for the apparently different scaling properties measured for some temperature range. Only further studies at increasing Reynolds numbers might be able to answer this question and clarify whether the differences between the two fluids and the apparently different scaling properties in the inertial range are Reynolds independent or not. On the other hand, the empirically observed enhancement of flatness for the Reynolds numbers investigated here is a robust observation, independent of the existence of any power law scaling.

A possible reason for lack of experimental evidence of a temperature dependence of turbulent statistics in coflowing $^4$He in \Ref{Roche-new} may  be a particular choice of the flow type, which is anisotropic and in which the turbulence is not fully developed at small scales, where the effect is observed.

\begin{acknowledgements}
We acknowledge funding from the Prace project "Superfluid Turbulence under counter-flows" Pra12\_ 3088.
LB and GS acknowledge funding from the European Union’s Seventh Framework Programme (No. FP7/2007-2013) under Grant Agreement No. 339032.

\end{acknowledgements}


\begin{thebibliography}{99}
\bibitem{Rev1} C. F. Barenghi, L. Skrbek, and K. R. Sreenivasan, \emph{Introduction to quantum turbulence},
Proc. Nat.  Acad. Sci. USA \textbf{111}, 4647-4652 (2014).
\bibitem{11} H. E. Hall and W. F. Vinen,
\emph{The Rotation of Liquid Helium II. I. Experiments on the
Propagation of Second Sound in Uniformly Rotating Helium II}, Proc. Roy. Soc. A
\textbf{238}, 204 (1956).
\bibitem{HV} H. E. Hall and W. F. Vinen, Proc. Roy. Soc. \emph{A The Rotation of Liquid Helium II. I. Experiments on the Propagation of Second Sound in Uniformly Rotating Helium II,} \textbf{238}, 204 (1956).
\bibitem{BK} I.L. Bekarevich, and I.M. Khalatnikov, \emph{Phenomenological Derivation of the Equations of Vortex Motion in He II,} Sov. Phys. JETP \textbf{13}, 643 (1961).

\bibitem{12} I.L. Bekarevich, and I.M. Khalatnikov,
 \emph{Phenomenological Derivation of the Equations of Vortex Motion in He II},
Sov. Phys. JETP \textbf{13}, 643 (1961).
 \bibitem{Tabeling}  J. Maurer and P. Tabeling, \emph{Local investigation of superfluid turbulence}, Europhys. Lett. \textbf{43}, 29 (1998).
 \bibitem{Roche1}  J. Salort, et al. \emph{Turbulent velocity spectra in superfluid flows}.
Phys Fluids \textbf{22}, 125102 (2010).
 \bibitem{Salort11} J. Salort, B. Chabaud, E. L\'{e}v\'{e}que, and P.E. Roche. \emph{Investigation of intermittency in superuid turbulence}. Jour. Phys. : Conf. Series, \textbf{318} (2011).
  \bibitem{Roche2} P.-E. Roche, C.F. Barenghi, E. L\'{e}v\'{e}que \emph{Quantum turbulence at finite
temperature: The two-fluids cascade}, Europhys Lett. \textbf{ 87}, 54006 (2009).
\bibitem{Rev2}C. F. Barenghi, V. S. L'vov, and P.-E. Roche,  \emph{Experimental, numerical, and analytical velocity spectra in turbulent quantum fluid}, Proc. Nat.  Acad. Sci. USA \textbf{111}, 4683-4690 (2014).
\bibitem{Rev3}V. Eltsov, R. Hanninen, M. Krusius, \emph{Quantum turbulence in superfluids with wall-clamped normal component}. Proc. Nat. Acad. Sci. USA \textbf{111}, 4711 (2014).
\bibitem{Roche-new}E. Rusaouen, B. Chabaud, J. Salort, Philippe-E. Roche.
\emph{Intermittency of quantum turbulence with superfluid fractions from 0\% to 96\%}. Submitted to Physics of Fluids in 2017.
\bibitem{WeiEmilGrid}Tallahassee group, private communication.
\bibitem{He4} L. Bou\'{e}, V.S. L'vov, A. Pomyalov and I. Procaccia, \emph{Enhancement of intermittency in superfluid turbulence}, Phys. Rev. Letts., \textbf{110}, 014502 (2013).
\bibitem{Shukla}V. Shukla and R. Pandit, \emph{Multiscaling in superfluid turbulence: A shell-model study}. Phys. Rev. E \textbf{94}, 043101(2016).
\bibitem{GPEgrid}G. Krstulovic, \emph{Grid superuid turbulence and intermit-tency at very low temperature.}
 Phys. Rev. E, \textbf{93}, 063104 (2016).

\bibitem{DB98} R. J. Donnelly and C.F.Barenghi,   \emph{The observed properties of liquid helium at the saturated vapor pressure}. J. Phys. Chem. Ref. Data\textbf{ 27}, 1217(1998).

\bibitem{DNS-He3} L. Biferale, D. Khomenko, V. L'vov, A. Pomyalov, I. Procaccia and G. Sahoo, \emph{Local and non-local energy spectra of superfuid $^3$He turbulence}, Phys. Rev. B \textbf{95}, 184510 (2017).
\bibitem{LNV} V. S. L'vov, S. V. Nazarenko and G. E. Volovik, \emph{Energy spectra of developed superfluid turbulence}, JETP Letters, \textbf{80},  535 (2004).
\bibitem{LP-QFS} V.S. L'vov, A. Pomyalov, \emph{Statistics of quantum turbulence in superfluid He}, JLTP  \textbf{187},  497 (2017).
\bibitem{LP-95} V.S. L'vov and I. Procaccia, I. \emph{Exact Resummations In The Theory Of Hydrodynamic Turbulence .1. The Ball Of Locality And Normal Scaling},  Physical Review E. \textbf{52}, 3840 (1995).
\bibitem{LP-2}  V.S. L'vov, I. Procaccia,  \emph{Exact Resummations In The Theory Of Hydrodynamic Turbulence .2. A ladder to anomalous  scaling},  Physical Review E. \textbf{52}, 3858 (1995).

\bibitem{45law}A.N. Kolmogorov, \emph{Dissipation of Energy in the Locally Isotropic Turbulence}, Dokl. Akad. Nauk. SSSR, \textbf{32}, 16 (1941).
\bibitem{LPP-97} V. S. L'vov , E. Podivilov  and I. Procaccia,  \emph{Exact Result for the 3rd Order Correlations of Velocity in Turbulence with Helicity}, arXiv:chao-dyn/9705016 v2.
    \bibitem{PRB} L. Bou\'{e},  V. S. L'vov,  Y. Nagar,  S. V. Nazarenko, A. Pomyalov,  and I. Procaccia, \emph{Energy and vorticity spectra in turbulent superfluid 4He from T = 0 to $T_{\lambda}$}, Phys.Rev.B \textbf{91}, 144501 (2015).
\bibitem{Frisch} U. Frisch, \emph{Turbulence: The Legacy of A. N. Kolmogorov}, (Cambridge University Press, 1995).

\bibitem{VN} W.F. Vinen, J.J. Niemela, \emph{Quantum turbulence}. J. Low Temp. Phys. 128, 167 (2002).
\bibitem{LNS-2006} V.S. L’vov, S.V. Nazarenko, L. Skrbek, \emph{Energy spectra of developed turbulence in helium superﬂuids}.
J. Low Temp. Phys. \textbf{145}, 125 (2006).


\bibitem{ess}R. Benzi, S. Ciliberto, R. Tripiccione, C. Baudet, F. Massaioli, and S. Succi, \emph{Extended self-similarity in turbulent flows}. Phys. Rev. E 48, R29(R)(1993).
\bibitem{SL}Z-S. She and E.Leveque, \emph{Universal scaling laws in fully developed turbulence}. Phys. Rev. Lett., \textbf{72}, 336 (1994).













\end{thebibliography}
\end{document}